\documentclass[12pt,preprint]{aastex}

\shortauthors{Watanabe et al.}

\begin{document}

\title{Solar Neutron Events of October-November 2003}

\author{K. Watanabe\altaffilmark{1}, M. Gros\altaffilmark{2}, P. H. Stoker\altaffilmark{3}, K. Kudela\altaffilmark{4}, C. Lopate\altaffilmark{5}, J. F. Vald\'{e}s-Galicia\altaffilmark{6}, \\ A. Hurtado\altaffilmark{6}, O. Musalem\altaffilmark{6}, R. Ogasawara\altaffilmark{7}, Y. Mizumoto\altaffilmark{7}, M. Nakagiri\altaffilmark{7}, A. Miyashita\altaffilmark{7}, \\ Y. Matsubara\altaffilmark{1}, T. Sako\altaffilmark{1}, Y. Muraki\altaffilmark{1}, T. Sakai\altaffilmark{8}, and S. Shibata\altaffilmark{9}}

\altaffiltext{1}{Solar-Terrestrial Environment Laboratory, Nagoya University, Nagoya, 464-8601, Japan}
\altaffiltext{2}{DSM/DAPNIA/SAp, CEA Saclay, 91191 Gif-sur-Yvette, France}
\altaffiltext{3}{Potchefstroom University, South Africa}
\altaffiltext{4}{Institute of Experimental Physics SAS, Kosice, Slovakia}
\altaffiltext{5}{The University of New Hampshire, Space Science Center, New Hampshire, 03824, USA}
\altaffiltext{6}{Instituto de Geophysica, Universidad National Aut\'{o}noma de M\'{e}xico, 04510, D. F. M\'{e}xico, M\'{e}xico}
\altaffiltext{7}{National Astronomical Observatory of Japan, Hilo, Hawaii, 96720, USA}
\altaffiltext{8}{College of Industrial Technologies, Nihon University, Narashino, 275-0005, Japan}
\altaffiltext{9}{College of Engineering, Chubu University, Kasugai, 487-8501, Japan}

\begin{abstract}
During the period when the Sun was intensely active on October-November 2003, two remarkable solar neutron events were observed by the ground-based neutron monitors.
On October 28, 2003, in association with an X17.2 large flare, solar neutrons were detected with high statistical significance ($6.4 \sigma$) by the neutron monitor at Tsumeb, Namibia.
On November 4, 2003, in association with an X28 class flare, relativistic solar neutrons were observed by the neutron monitors at Haleakala in Hawaii and Mexico City, and by the solar neutron telescope at Mauna Kea in Hawaii simultaneously.
Clear excesses were observed at the same time by these detectors, with the significance calculated as $ 7.5\,\sigma $ for Haleakala, and $ 5.2\,\sigma $ for Mexico City.
The detector onboard the {\it INTEGRAL} satellite observed a high flux of hard X-rays and $\gamma$-rays at the same time in these events.
By using the time profiles of the $\gamma$-ray lines, we can explain the time profile of the neutron monitor.
It appears that neutrons were produced at the same time as the $\gamma$-ray emission.
\end{abstract}


\section{Introduction}
\label{introduction}

Relativistic particles, in particular solar neutrons, give information about ion acceleration in solar flares.
Several observations of solar neutrons in solar cycle 23 have been reported using the international network of neutron monitors \citep{Usoskin1997, LockwoodDebrunner1999} and solar neutron telescopes \citep{Tsuchiya2001, Valdes2004}.
Solar flares which produce neutrons frequently occur when activity is near its maximum during a solar cycle.
In most cases they have also produced X-class solar flares.
More than one hundred X class flares have been recorded in this solar cycle.

Intense solar activity occurred from late October to the beginning of November 2003.
The events that occurred in this period were observed by numerous satellites and detectors and have been analyzed by many investigators.
During the period when three active regions appeared simultaneously on the Sun the soft X-ray flux was very intense and a series of eleven X-class solar flares occurred in NOAA regions 10484, 10486 and 10488.
At this time, solar neutrons were observed on October 28 and November 4, 2003, in association with X17.2 and X28 class solar flares, respectively.

An X17.2 solar flare on October 28, 2003, was a remarkable event in this solar cycle.
Not only was this a large event, but many phenomena were observed in association with this flare.
It is particularly worth noting the large flux of relativistic particles at the Earth \citep{Veselovsky2004, Panasyuk2004}.
Among these particles were solar neutrons which were observed by the ground-based neutron monitor before the main Ground Level Enhancement (GLE).
This solar neutron event has already been reported and discussed by \citet{Bieber2005}.
In this paper, we compare neutron data with $\gamma$-ray data observed by {\it INTEGRAL} satellite and derive the energy spectrum of neutrons using these $\gamma$-ray data.

The second event occurred on November 4, 2003, and solar neutrons were observed by NM64 type neutron monitors located at different places, one at Haleakala in Hawaii and the other at Mexico City in Mexico.
In addition, solar neutrons were also observed by a solar neutron telescope located at Mauna Kea in Hawaii.
Thus, it is important that a single model can explain the data of three detectors to lead an accurate spectrum of solar neutrons from the solar neutron event.

Simultaneous observations of solar neutrons have been made for a few events.
In the solar event of June 3, 1982, neutrons were observed by a ground-level detector and by spacecraft, simultaneously \citep{Chupp1987}.
High energy neutrons were detected by the IGY type neutron monitor installed at Jungfraujoch, Switzerland, and low energy neutrons and high energy $\gamma$-rays were observed by the Gamma Ray Spectrometer (GRS) onboard the {\it Solar Maximum Mission (SMM)}.

On May 24, 1990, solar neutrons were observed by the IGY type neutron monitors located at Climax and several stations in North America, simultaneously \citep{Shea1991, Debrunner1997, MurakiShibata1996}.
On June 4, 1991, solar neutron signals were recorded by the neutron monitor and the solar neutron telescope located at Mt.\,Norikura \citep{Muraki1992, Struminsky1994}.
However, in this event, the energy spectrum of solar neutrons calculated from the data of these detectors were not self consistent.
This discrepancy came from the propagation model of solar neutrons in the Earth's atmosphere.
By using the same propagation model, which is Shibata model \citep{Shibata1994}, nearly the same spectrum was obtained \citep{Shibata1993}.
In this paper, we report the analysis results of the November 4 event using data from the neutron monitors, the solar neutron telescope and the spacecraft.
Our model can explain data from three detectors consistently.

\section{Solar Neutron Event Associated with an X17.2 Flare on 2003 October 28}

\subsection{Observations}
\label{observations_20031028}
\clearpage
\begin{figure}[tbp]
    \includegraphics[scale=0.65]{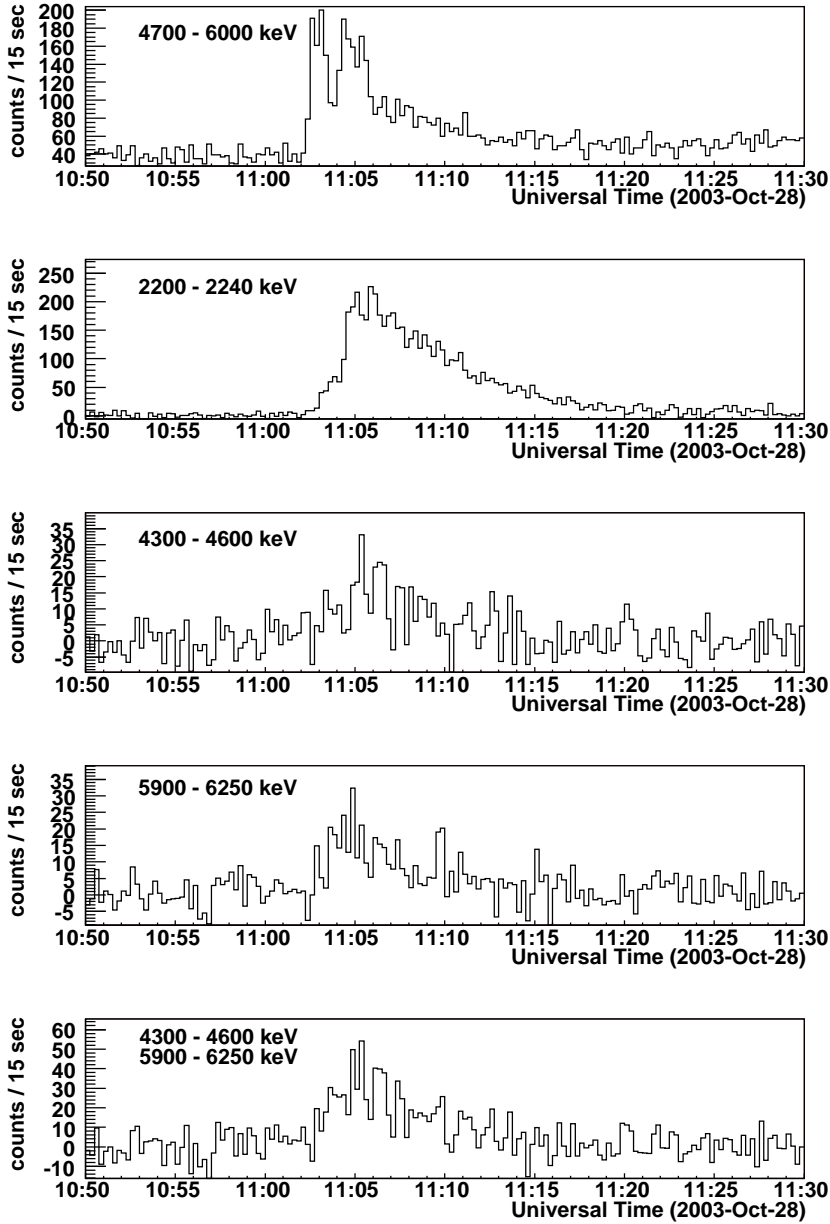}
  \caption{Top panel shows the time profiles of bremsstrahlung $\gamma$-rays observed by the {\it INTEGRAL} satellite on October 28, 2003. Second to fourth panel shows the line $\gamma$-ray time profiles observed by the {\it INTEGRAL} satellite on October 28, 2003. The bremsstrahlung component has been subtracted. Second panel shows the time profile of the $ 2.2\,{\rm MeV} $ neutron capture $\gamma$-rays. Third panel gives the profile of $ 4.4\,{\rm MeV} $ $\gamma$-ray of C nuclei, and the fourth one that of the $ 6.1\,{\rm MeV} $ $\gamma$-ray of O nuclei. The final panel shows the sum of the data in the third and fourth panels.}
  \label{20031028_integral}
\end{figure}

\begin{figure}[tbp]
    \includegraphics[scale=0.40]{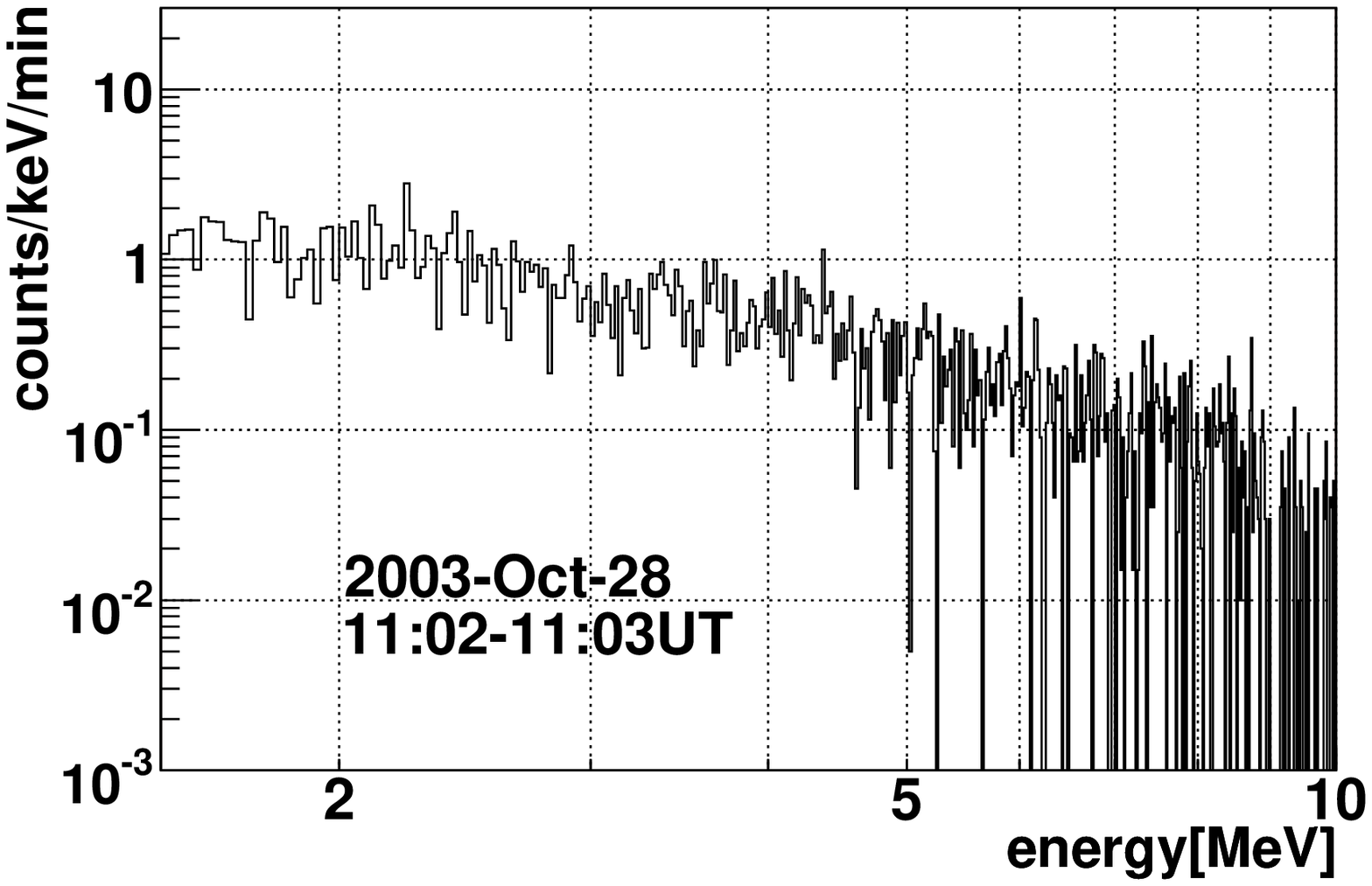}
    \includegraphics[scale=0.40]{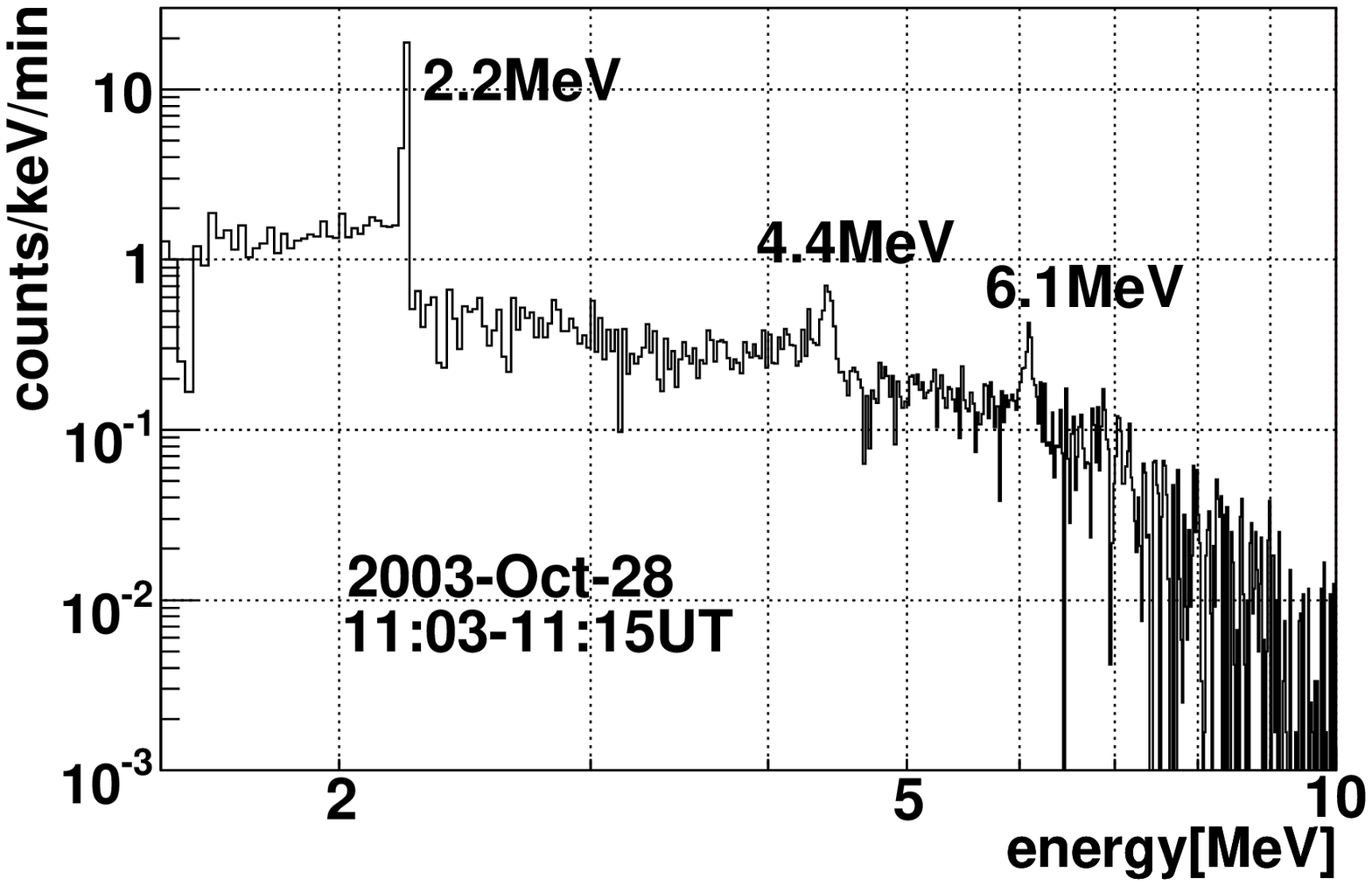}
  \caption{Spectra of $\gamma$-rays between $1.5$ and $10\,{\rm MeV}$ observed by {\it INTEGRAL} from 11:02$-$03\,UT (left) and 11:03$-$15\,UT (right) on October 28, 2003, with background subtracted. In right figure, clear signals of $ 2.2 $, $ 4.4 $ and $ 6.1\,{\rm MeV} $ $\gamma$-rays appear superimposed on the bremsstrahlung component.}
  \label{20031028_integral_spec}
\end{figure}
\clearpage
An X17.2 class solar flare occurred at 9:51\,UT (this is the time observed at the Earth, and same definition here after) on October 28, 2003, located in NOAA active region 10486 at $ {\rm S} 16 ^{\circ} $ $ {\rm E} 08 ^{\circ} $.
From 10:36 to 11:06\,UT, an interval which includes the start time of intense emission of soft X-rays of the X17.2 flare, the {\it RHESSI} satellite was, unfortunately, in the South Atlantic Anomaly (SAA).
However, intense emission of high energy $\gamma$-rays was seen in the data after 11:06\,UT, indicating that strong particle acceleration occurred during this flare.

On the other hand, large fluxes of hard X-rays and $\gamma$-rays were observed by the {\it INTEGRAL} satellite shortly after 11:00\,UT.
Figure \ref{20031028_integral} show the bremsstrahlung and line $\gamma$-ray time profiles from {\it INTEGRAL}.
In the top panel of Figure \ref{20031028_integral}, two peaks of intense emission of bremsstrahlung $\gamma$-rays are seen at around 11:03 and 11:05\,UT.
However, there is only one peak (around 11:05\,UT) in line $\gamma$-ray time profiles as shown in the second to fourth panel in Figure \ref{20031028_integral}.
This more or less coincides with the second peak in the bremsstrahlung $\gamma$-rays.

Figure \ref{20031028_integral_spec} shows $\gamma$-ray spectra between 11:02 and 11:03\,UT, and between 11:03 and 11:15\,UT.
From 11:02 to 11:03\,UT, when the first peak of bremsstrahlung $\gamma$-rays was seen, there is no line $\gamma$-ray component.
$\gamma$-ray lines were clearly seen in the $\gamma$-ray spectrum after 11:03\,UT, consistent with the line $\gamma$-ray time profiles shown in Figure \ref{20031028_integral}.
Thus, it appears that the time profile of ion and electron acceleration were quite different at this event, and ion acceleration either didn't occur or was quite weak during the first peak of the bremsstrahlung $\gamma$-rays.
We can assume that the ion acceleration only started after 11:03\,UT.

In Figure \ref{20031028_integral}, note that the $ 2.2\,{\rm MeV} $ neutron capture $\gamma$-ray line peaks around 11:06\,UT and has a long decay time.
The $ 4.4 $ and $ 6.1\,{\rm MeV} $ $\gamma$-ray lines of de-excited $ {\rm C} $ and $ {\rm O} $ ions peak around 11:05\,UT, giving about a one minute gap between the two peak times.
In general, neutron capture $\gamma$-rays are delayed from $\gamma$-ray lines of de-excited ions, since it takes time for high energy neutrons to slow down and be captured by protons \citep{WangRamaty1974}.
Thus, it is evident that solar neutrons were produced at this flare, and were probably produced at the same time that $ 4.4 $ and $ 6.1\,{\rm MeV} $ $\gamma$-ray lines were emitted.
Hereafter, we assume that solar neutrons were produced around 11:05\,UT.

At 11:05\,UT on October 28, 2003, the Sun was located over Africa.
Among our international network of solar neutron telescopes, Gornergrat in Switzerland and Aragats in Armenia had a possibility of observing solar neutrons.
On the other hand, Tsumeb observatory ($17.6^{\circ}$E, $19.1^{\circ}$S, $1240\,{\rm m}$ a.s.l.) was located just under the Sun at this time.
The altitude of Tsumeb Observatory is a little bit low, however the air mass for the line of sight to the Sun was thinner than any other observatory because the zenith angle of the Sun was $ 9.5 ^{\circ} $.
Solar neutrons were clearly observed by the Tsumeb neutron monitor \citep{Bieber2005}.
\clearpage
\begin{figure}[tbp]
    \includegraphics[scale=0.80]{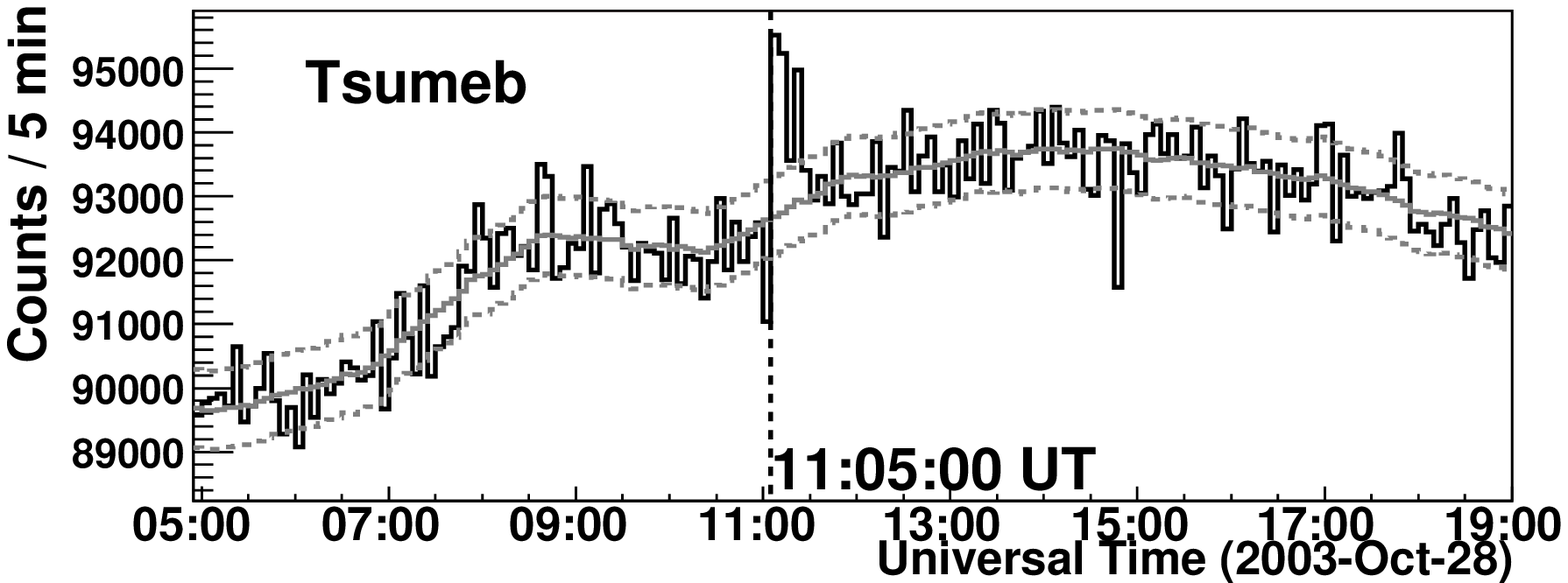}
    \includegraphics[scale=0.80]{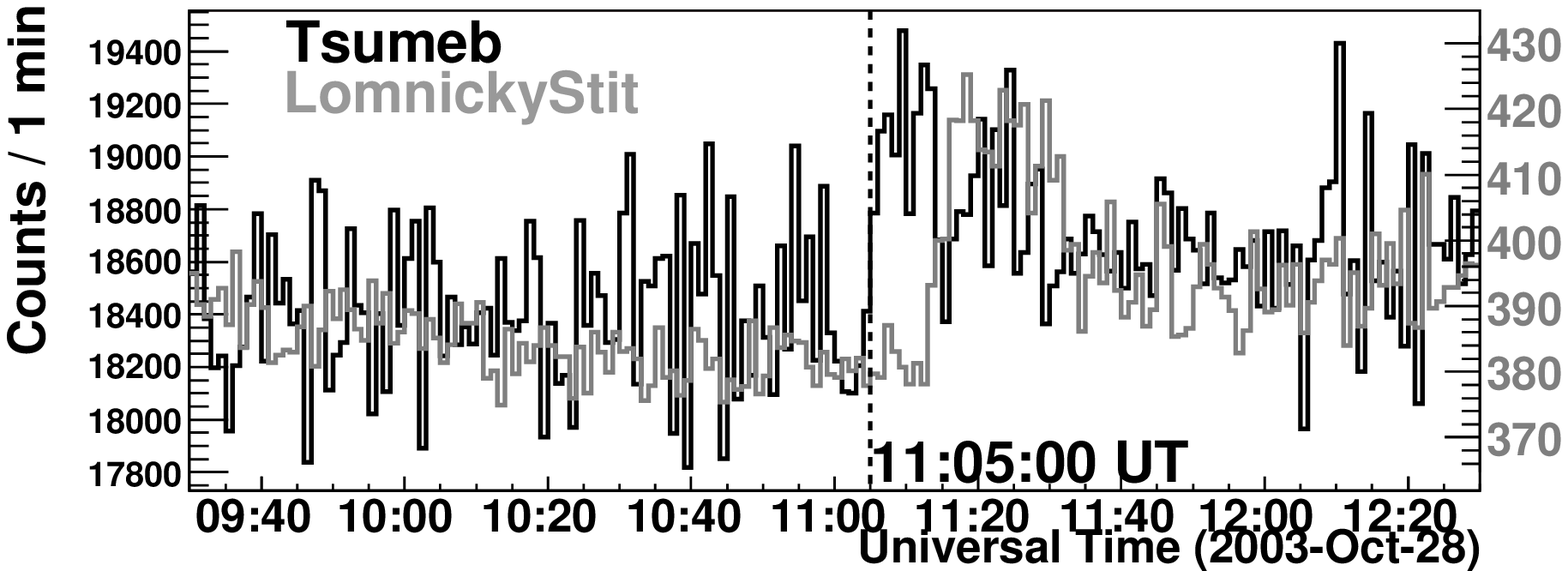}
  \caption{Top panel shows the five minute counting rate observed by the Tsumeb neutron monitor on October 28, 2003. The solid smooth line is the averaged background and the dashed lines are $\pm 1 \sigma$ from the background. The bottom panel gives the one minute counting rate of the Tsumeb neutron monitor (black line) and the time profile of the Lomnicky Stit neutron monitor (gray line). The solar neutron event in the Tsumeb data started well before the GLE event seen at Lomnicky Stit.}
  \label{20031028_NM}
\end{figure}
\clearpage
The five minute counting rate of the Tsumeb neutron monitor is shown in top panel of Figure \ref{20031028_NM}.
Clear excesses are seen between 11:05 and 11:15\,UT and between 11:20 and 11:25\,UT.
The statistical significance of these excesses are $ 4.8 \,\sigma $ for 11:05$-$11:10\,UT, $ 4.2 \,\sigma $ for 11:10 $-$ 11:15\,UT and $ 3.4 \,\sigma $ for 11:20$-$11:25\,UT.
The total significance for the ten minutes between 11:05\,UT and 11:15\,UT is $ 6.4 \,\sigma $.

At the same time, high energy protons were produced in association with this flare, and large ground level enhancements (GLEs) occurred around the world.
We can exclude the possibility that excesses observed at Tsumeb came from energetic ions, by considering the time profile of the neutron monitor at Lomnicky Stit ($20.2^{\circ}$E, $49.2^{\circ}$N, $2634\,{\rm m}$ a.s.l.) together with that of the Tsumeb's neutron monitor (bottom panel of Figure \ref{20031028_NM}).
The start time of the first excess of the Tsumeb neutron monitor is about $10$ minutes earlier than the event at Lomnicky Stit, while the second excess at Tsumeb is consistent with this event at Lomnicky Stit.
Thus, it appears that the second excess at Tsumeb came from energetic ions and the first excess was solar neutrons.

\subsection{Analysis result}
\label{analysis_20031028}

Using observational data presented in Section \ref{observations_20031028}, we can calculate the energy spectrum of the solar neutrons even though neutron monitors cannot measure the energy of neutrons.
By using the time of flight (TOF) method and assuming the emission time of solar neutrons, a spectrum can be derived. 
We assume that the neutrons were produced at 11:05\,UT, when line $\gamma$-ray emission peaked and that the energy of the neutrons responsible for the excesses recorded by the neutron monitor is greater than $ 100\,{\rm MeV} $.

From the time profile of the neutrons, we calculate the energy spectrum of solar neutrons at the solar surface by the following formula:
\begin{equation}
\frac{ \Delta N }{ \epsilon \times S \times \Delta E_n \times P}
\label{eq:flux}
\end{equation}
where $ \Delta N $ is the number of excess counts contributed by solar neutrons, and $ \epsilon $ is the detection efficiency of the neutron monitor.
Here, $ \epsilon $ includes the attenuation of solar neutrons through the Earth's atmosphere.
$ S $ is the area of the neutron monitor, $ \Delta E_n $ is the energy range corresponding to one time bin and $ P $ is the survival probability of solar neutrons traveling from the Sun to the Earth.

To obtain the $ \epsilon $ of Equation (\ref{eq:flux}), we calculated the attenuation of solar neutrons by the Earth's atmosphere using the Shibata model \citep{Shibata1994}.
Solar neutrons with energy less than $ 100 \,{\rm MeV} $ are strongly attenuated by the Earth's atmosphere, so the detection of neutrons by the Tsumeb monitor directly implies that the spectrum extended beyond $ 100 \,{\rm MeV} $.
For the detection efficiency of the neutron monitor, we used calculated result by \citet{ClemDorman2000}.
\clearpage
\begin{figure}[tbp]
    \includegraphics[scale=0.80]{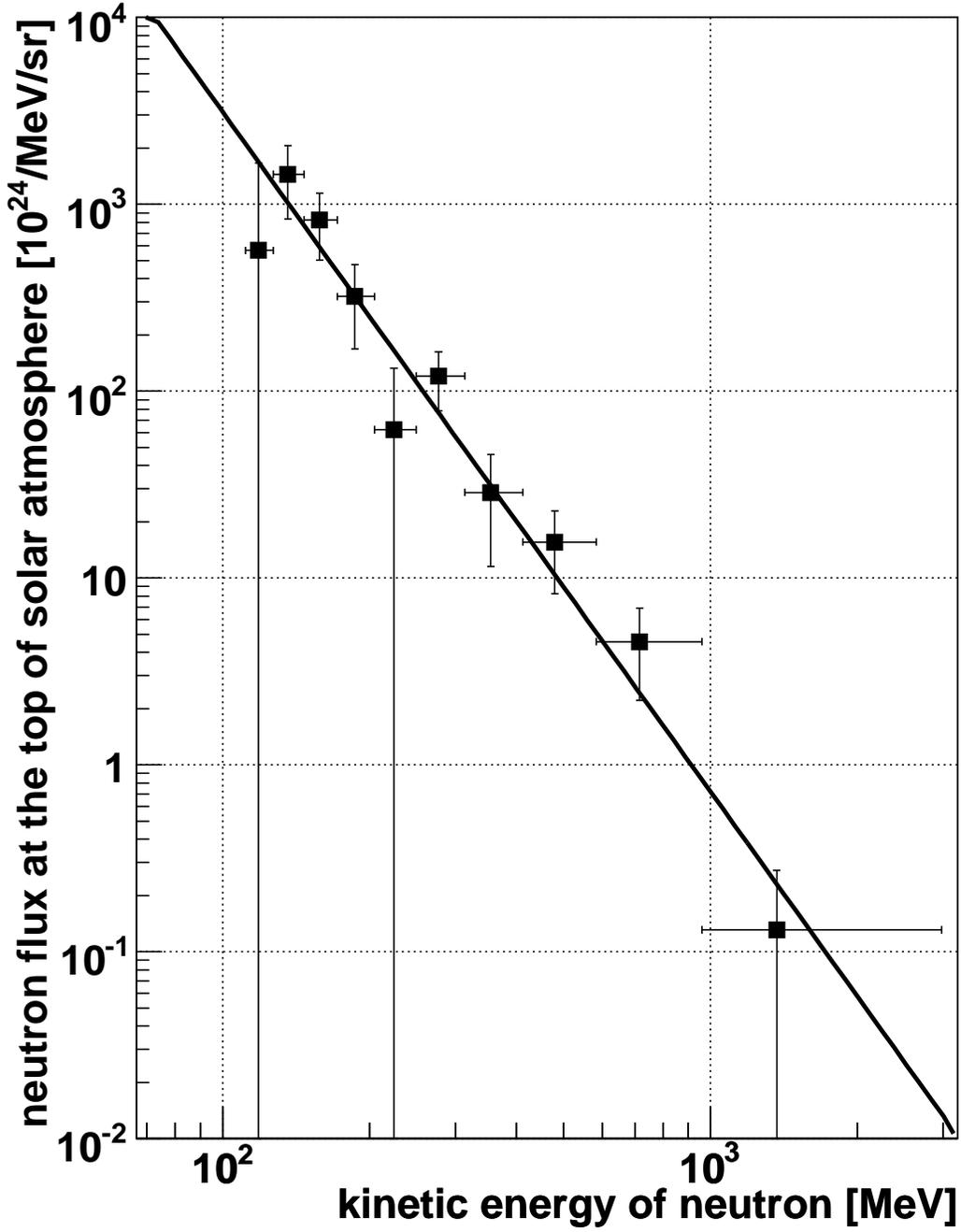}
  \caption{The energy spectrum of neutrons at the solar surface on October 28, 2003.}
  \label{20031028_energy_spec}
\end{figure}
\clearpage
Using these observational and simulation results, we calculated the energy spectrum of neutrons at the solar surface using the method used for the solar neutron event observed on November 24, 2000 \citep{Watanabe2003}.
Figure \ref{20031028_energy_spec} shows the result from Equation (\ref{eq:flux}).
By fitting these data points with a power law of the form $ C \times (E_n/100[{\rm MeV}])^{\alpha} $, the energy spectrum of solar neutrons was obtained.
The energy spectrum is fitted by a power law as;
\begin{eqnarray}
(3.1 \pm 1.0) \times 10^{27} \times \left( \frac{E_n}{100\,{\rm [MeV]}} \right) ^{-3.6 \pm 0.3}\,{\rm [/MeV/sr]}.
\label{eq:flux_20031028}
\end{eqnarray}
For this fit, $ \chi ^2 / {\rm dof} = 7.10 / 8 = 0.89 $, and the $ \chi ^2 $ probability is 53 \%.
The fitting region is above $ 100\,{\rm MeV} $.
This power index is a typical value for solar neutron events observed thus far.
The total energy flux of $ > 100\,{\rm MeV} $ neutrons emitted by the Sun was estimated to be $ 3.1 \times 10^{25}\,{\rm erg/sr} $.

\subsubsection{Simulation by Impulsive Model}
\label{sim_delta_20031028}

In the analysis method described above, the energy spectrum is calculated by dividing the response into several bins, each characterized by a mean energy.
For the survival probability of solar neutrons, as well as the attenuation of neutrons and detection efficiency of the neutron monitor, the values at these discrete energies are used.
In order to calculate the energy spectrum of the solar neutrons in detail, we include an assumption about the time profiles of solar neutrons but still assume a power law spectral index at the solar surface.
Using this method, we can investigate whether the neutrons are produced continuously.
To clarify the consistency with the conventional method, we begin by assuming that the neutrons are produced impulsively.
\clearpage
\begin{figure}[tbp]
    \includegraphics[scale=0.75]{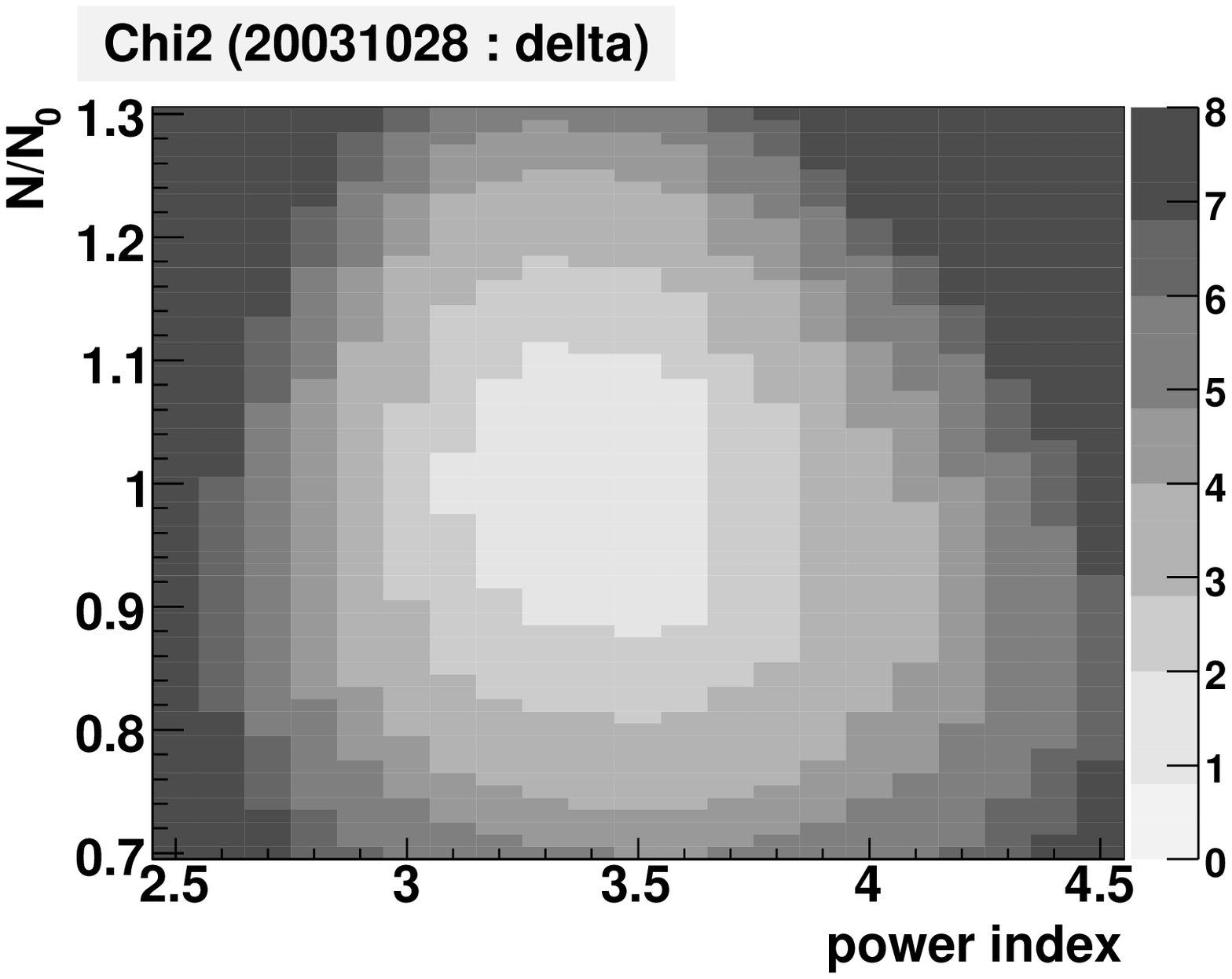}\\
    \includegraphics[scale=0.75]{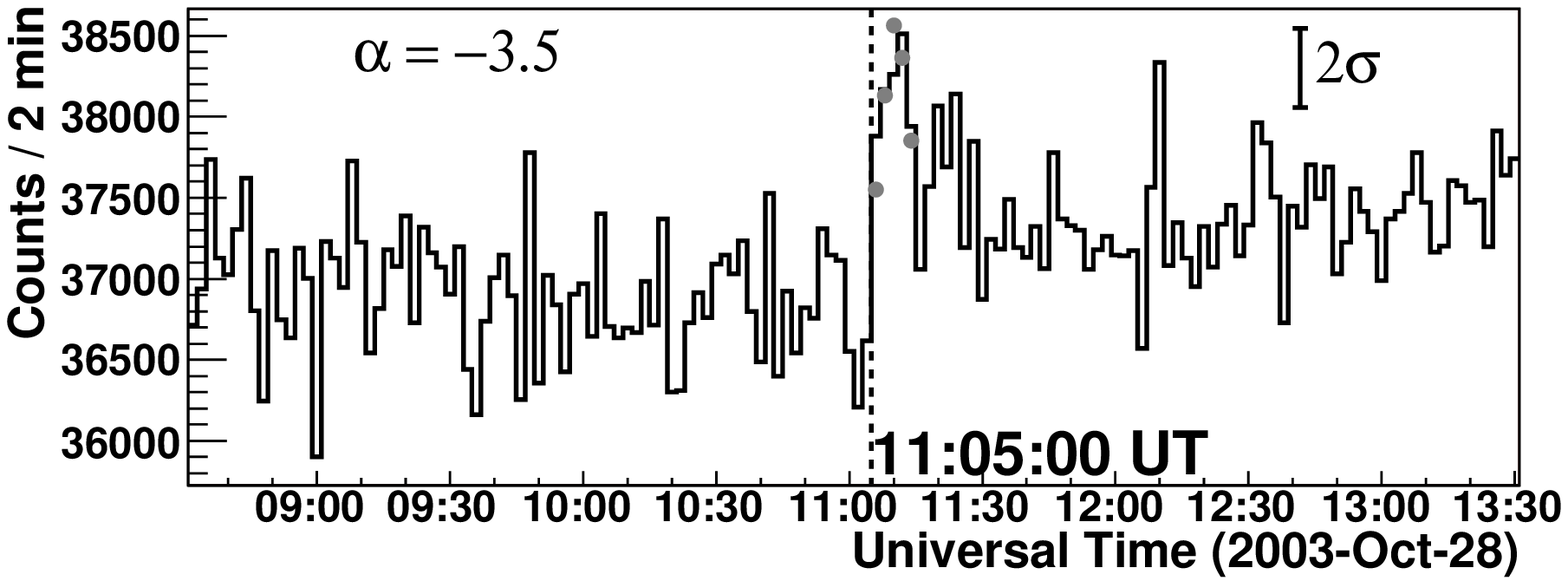}
  \caption{Top panel shows the reduced $ \chi ^2 $-distribution of the fit of the simulated counting rate to the observed excess of the Tsumeb neutron monitor. Two minute counting rate is used in this calculation. The X-axis represents the power index of the simulated time profiles, and Y-axis corresponds to the ratio of simulated counting rate ($N$) to the observed one ($N_0$). In this fitting, data obtained during 11:05$-$11:15\,UT are used. Bottom panel shows the two minute counting rate (solid line) observed by the Tsumeb neutron monitor on October 28, 2003, together with simulated time profile (points) for which solar neutrons are assumed to be produced instantaneously at 11:05\,UT, when power index is $-3.5$.}
  \label{20031028_sim_tprof}
\end{figure}
\clearpage
In this simulation, the power index of the neutron spectrum at the Sun is changed from $ -1.5 $ to $ -7.0 $ with a step of $ 0.1 $, while the energy range of the incident neutrons confined to $ 50-1500 {\rm \,MeV} $.
The time profile of neutrons detected by the neutron monitor is calculated using the neutron attenuation in the Earth's atmosphere given by the Shibata model \citep{Shibata1994} and the detection efficiency of the neutron monitor as calculated by \citet{ClemDorman2000}.
The decay of neutrons between the Sun and the Earth is also taken into account.
The result of this simulation can then be compared with the observational data, normalizing the simulated counting rate ($N$) to the observed excess counting rate ($N_0$).

The top panel of Figure \ref{20031028_sim_tprof} shows the reduced $ \chi ^2 $-distribution of the fit of the simulated counting rate to the observed excess of the Tsumeb neutron monitor obtained from the following formula:
\begin{eqnarray}
  \chi ^2 = \sum_{i=1}^{n} \frac{(N_{i} - N_{0i})^2}{N_{0i}}.
  \label{eq:chi2}
\end{eqnarray}
In this fitting, data obtained from 11:05 to 11:15\,UT are used.
$ \chi ^2 $ has its smallest value when the spectral index is  around $-3.5$.
When the spectral index is $-3.5$, the simulated result reproduces the observed result as shown in the bottom panel of Figure \ref{20031028_sim_tprof}, where $ \chi ^2 / {\rm dof} = 6.07 / 4 = 1.52 $, which yields the minimum $ \chi ^2 $ for the simulated time profile.
From this fitting, the energy spectrum is determined as follows:
\begin{eqnarray}
(3.3 \pm 0.3) \times 10^{27} \times \left( \frac{E_n}{100\,{\rm [MeV]}} \right) ^{-3.5 ^{+0.4} _{-0.2}}\,{\rm [/MeV/sr]}.
\label{eq:flux_20031028_delta}
\end{eqnarray}
This is comparable to the result obtained using the simpler method shown in Equation (\ref{eq:flux_20031028}), but the total energy flux of solar neutrons with energy range between $ 50 - 1500\,{\rm MeV} $ is $ (9.8 ^{+1.2} _{-0.9}) \times 10^{25}\,{\rm erg/sr} $, about a factor of three higher than the other estimate.
This is because of the lower cutoff energy of the neutron spectra.

\subsubsection{Simulation by Neutron Production with $\gamma$-ray Time Profile}
\label{sim_gamma_20031028}
\clearpage
\begin{figure}[tbp]
    \includegraphics[scale=0.75]{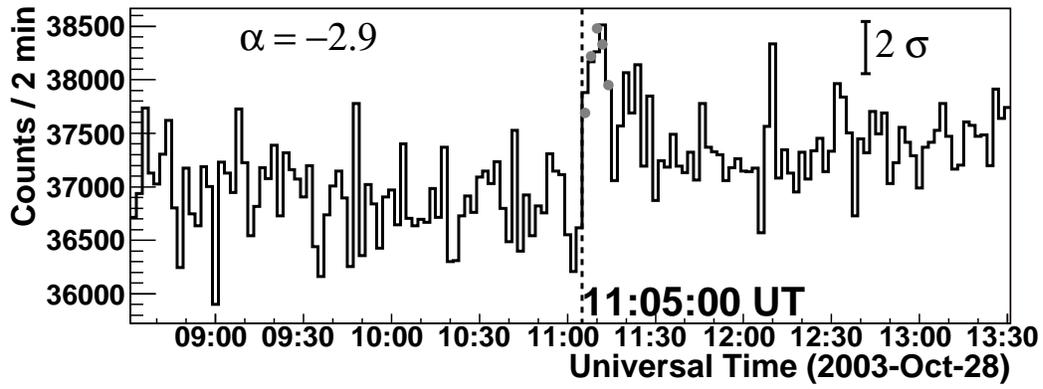}
  \caption{The observed and simulated time profiles of the Tsumeb neutron monitor. Solid line is the observed two minute counting rate and points indicate the best fit simulated time profile for solar neutrons assumed to be produced with the same time profile as $\gamma$-ray lines.}
  \label{20031028_sim_tprof_gamma}
\end{figure}
\clearpage
We next simulated the neutron time profiles detected at Tsumeb by assuming that neutrons were produced with a time spread, since extended production of line $\gamma$-ray was observed by {\it INTEGRAL}.
For this calculation, we used the summed time profile of $ 4.4 $ and $ 6.1\,{\rm MeV} $ $\gamma$-ray as shown in the bottom panel of Figure \ref{20031028_integral}, as the production model of solar neutrons.
These are the $\gamma$-ray lines of Carbon and Oxygen, which indicate the time profile of ion acceleration.
We used the data observed from 11:02:45 to 11:10:00\,UT.
The spectral index of neutrons at the Sun is varied from $ -1.1 $ to $ -7.0 $ in steps of $ 0.1 $.
The energy range of the neutrons is again taken to be $ 50 - 1500 {\rm \,MeV} $.

The $ \chi ^2 $ for the fit were calculated by using data obtained from 11:05 to 11:15\,UT.
$ \chi ^2 $ has its smallest value for the spectral index $-2.9$.
The simulated result reproduces the observed result most closely when the spectral index is $-2.9$  as shown in Figure \ref{20031028_sim_tprof_gamma}.
When the spectral index is $-2.9$, $ \chi ^2 / {\rm dof} =  2.76 / 4 = 0.69 $, which provides the minimum $ \chi ^2 $ for the simulated time profiles.
From this fit, the spectral index is found to be $ -2.9 \pm 0.3 $.
The best fit spectral index is harder than the index derived on the assumption that the  neutrons were produced impulsively, but the total energy flux of the neutrons is now estimated to be $ (6.2 ^{+0.5} _{-0.6}) \times 10^{25}\,{\rm erg/sr} $, not very much different from the result for impulsive production.

\section{Simultaneous Observations of Solar Neutrons on 2003 November 4}

\subsection{Observations}
\label{observations_20031104}
\clearpage
\begin{figure}[tbp]
    \includegraphics[scale=0.8]{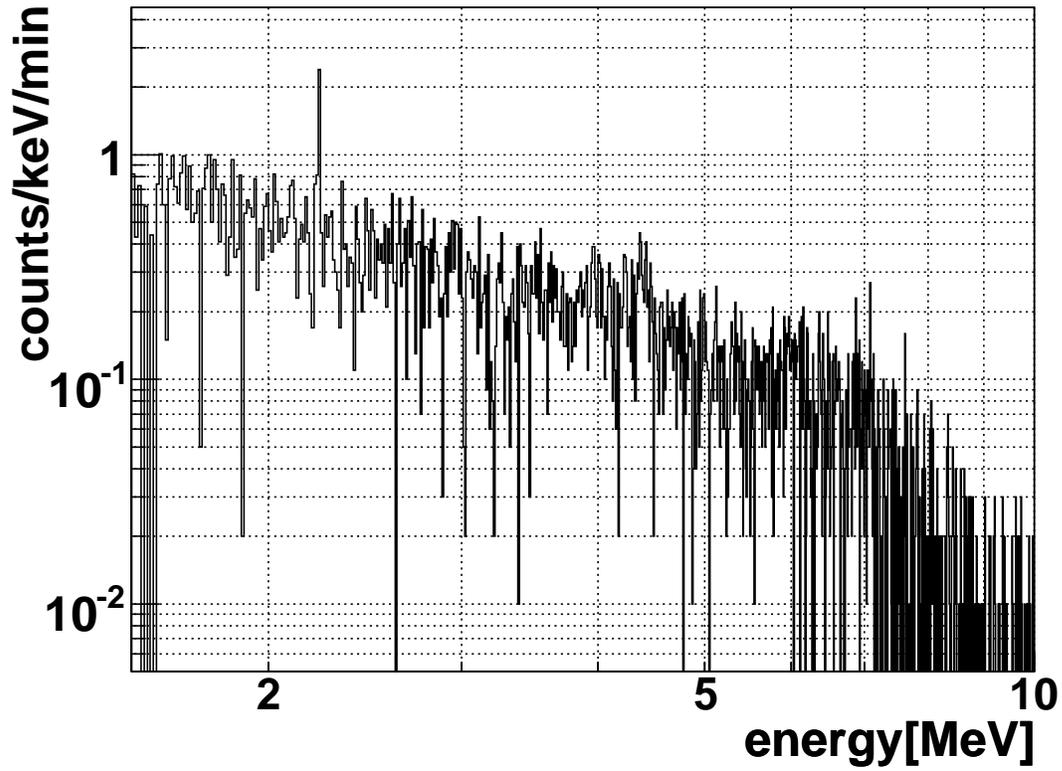}
  \caption{Spectrum of $\gamma$-rays between $1.5$ and $10\,{\rm MeV}$ observed by {\it INTEGRAL} from 19:40 to 19:50\,UT on November 4, 2003, with background subtracted. A signal produced by $ 2.2\,{\rm MeV} $ $\gamma$-rays appear superimposed on the bremsstrahlung component. There is weak evidence for $ 4-7\,{\rm MeV} $ $\gamma$-ray lines.}
  \label{20031104_integral_spec}
\end{figure}

\begin{figure}[tbp]
    \includegraphics[scale=0.8]{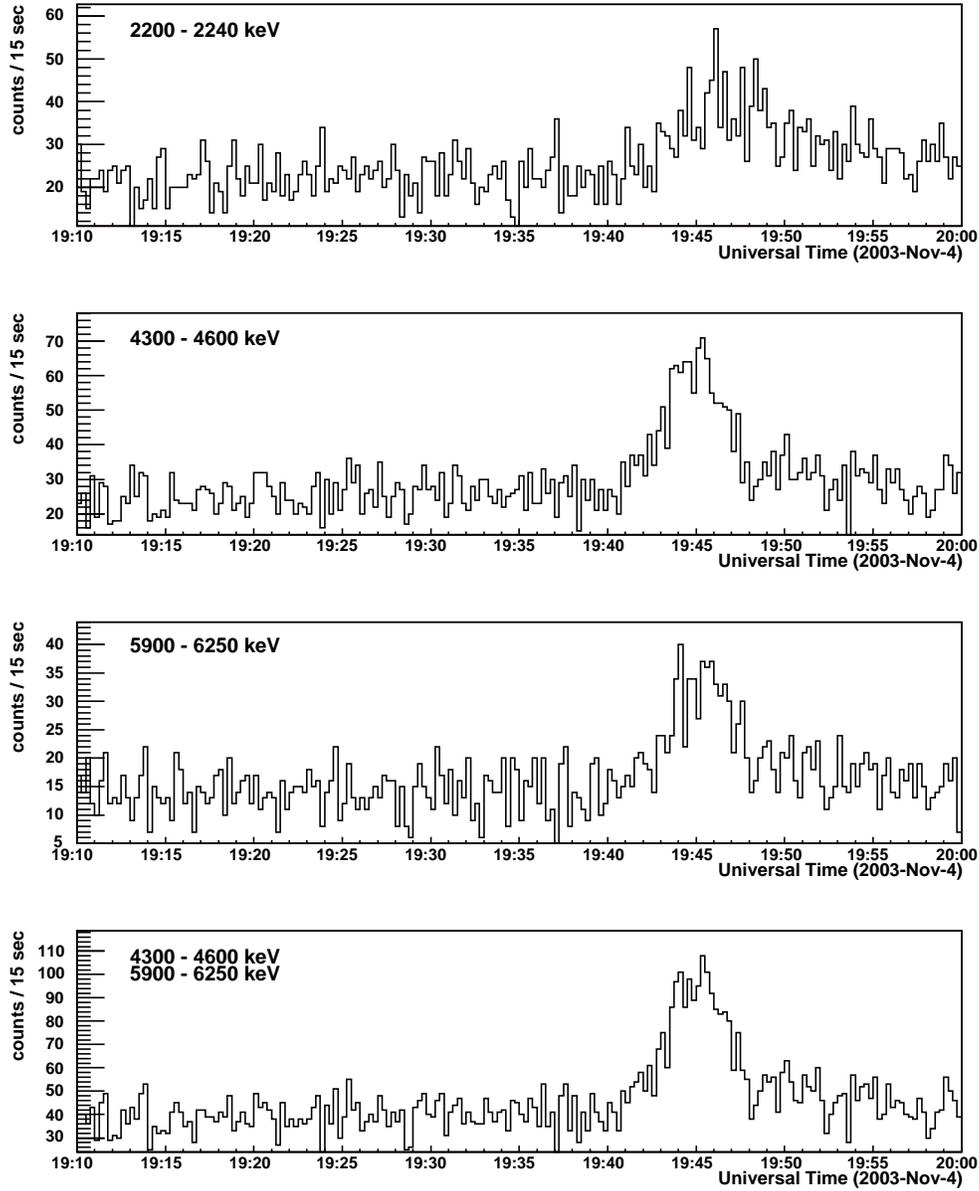}
  \caption{Time profiles of $\gamma$-ray lines observed by the {\it INTEGRAL} satellite on November 4, 2003. The bremsstrahlung component has not been subtracted. Top panel shows the time profile of the $ 2.2\,{\rm MeV} $ neutron capture $\gamma$-ray. Second panel gives the $ 4.4\,{\rm MeV} $ $\gamma$-ray of C nuclei, and the third one contains the $ 6.1\,{\rm MeV} $ $\gamma$-ray of O nuclei. The final panel shows the sum of the data in the second and third panels. Although these time profiles contain line $\gamma$-ray components which indicate the time profile of ion acceleration, the dominant component is bremsstrahlung.}
  \label{20031104_integral_line}
\end{figure}
\clearpage
On November 4, 2003, an X28 class solar flare occurred at 19:29\,UT, located in NOAA active region 10486 at $ {\rm S} 19 ^{\circ} $ $ {\rm W} 83 ^{\circ} $.
This is the largest solar flare on record.
At around 19:42\,UT, intense emission of soft X-rays was detected by the {\it GOES} satellite such that the detection was saturated.
After 19:42\,UT, intense emission of hard X-rays and $\gamma$-rays was observed by the {\it INTEGRAL} spacecraft.
Unfortunately, at this time, the {\it RHESSI} spacecraft was on the night side of the Earth.
Figure \ref{20031104_integral_spec} shows the energy spectrum of $\gamma$-rays observed by {\it INTEGRAL}.
In this event, although the components of the line emission produced by de-excited ions, C($ 4.4\,{\rm MeV} $) and O($ 6.1\,{\rm MeV} $) were not prominent, the $ 2.2\,{\rm MeV} $ neutron capture line can be clearly seen.
Intense bremsstrahlung X-rays and $\gamma$-rays were also observed.
Figure \ref{20031104_integral_line} shows the time profiles of $\gamma$-rays for different energy bins that contain line $\gamma$-ray components produced as a result of the ion acceleration.
There is a delay of the $ 2.2\,{\rm MeV} $ neutron capture $\gamma$-ray emission from that of the line $\gamma$-ray components produced by excited ions of C and O.
We may assume that ion acceleration occurred at the same time as the $\gamma$-ray lines were emitted, although the main component of these $\gamma$-rays is bremsstrahlung.
And we may assume that solar neutrons were produced at the same time.

At 19:45\,UT, the Sun was located between Hawaii and South America.
Although the Chacaltaya observatory was the best place to observe solar neutrons in our international solar neutron telescope network at this time, no data are available because of a data gap.
Sierra Negra would also have been a good place to observe solar neutrons, but the Mexico solar neutron telescope had not started continuous observation at that time.
Thus it was necessary to examine data from the Hawaii observatory, which was the third closest of the observatories able to observe solar neutrons.

At 19:45\,UT, the zenith angle of the Sun was $49.9 ^{\circ}$ at Mauna Kea, and $50.5 ^{\circ}$ at Haleakala.
The air mass along the line of sight to the Sun was $947\,\rm{g/cm^2}$ and $1112\,\rm{g/cm^2}$, respectively.
The other suitable location was Mexico City, where the zenith angle of the Sun was $40.52 ^{\circ}$ and the air mass along the line of sight to the Sun was $1026\,\rm{g/cm^2}$.
Attenuation of solar neutrons by the Earth's atmosphere above these observatories is calculated using the Shibata model \citep{Shibata1994}.
The Haleakala and the Mexico City observatories have nearly the same attenuation, while Mauna Kea is located at the best place to observe solar neutrons.
However, simultaneous signals were found in both the Haleakala and the Mexico City neutron monitors.
\clearpage
\begin{figure}[tbp]
    \includegraphics[scale=0.8]{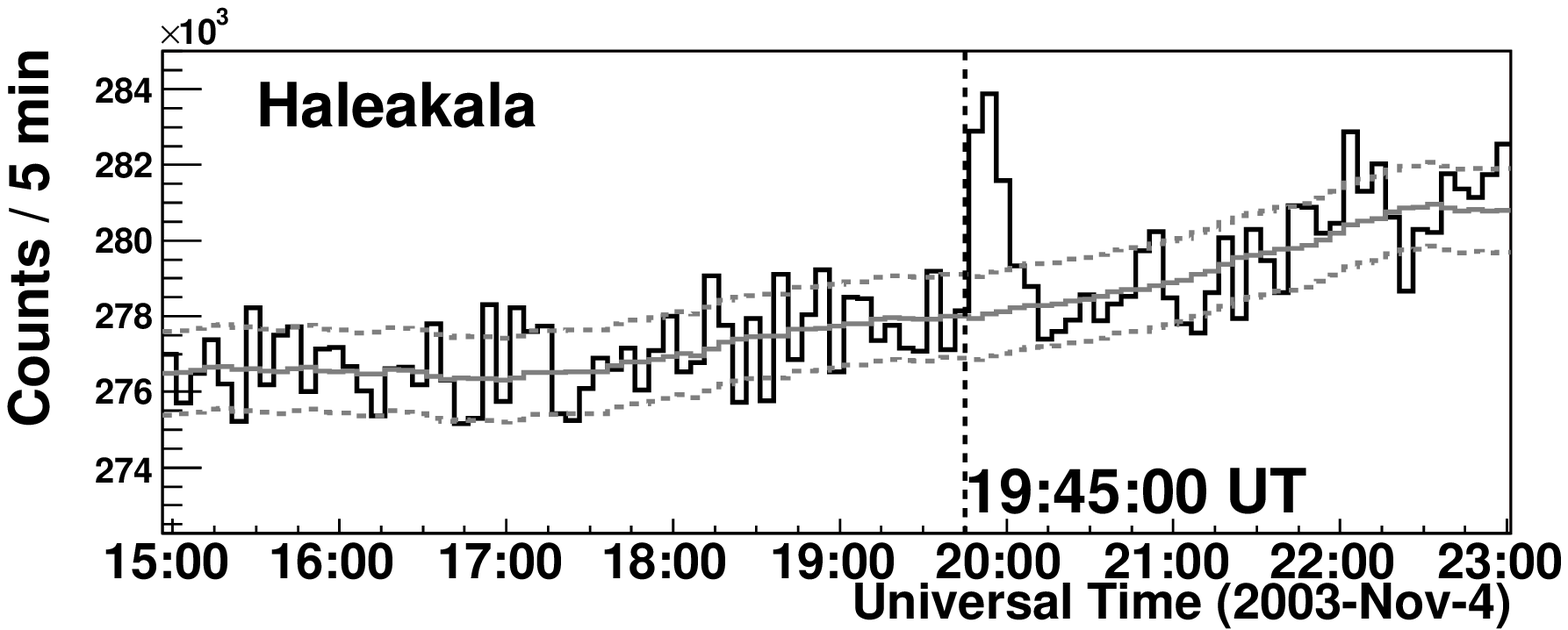}
    \includegraphics[scale=0.8]{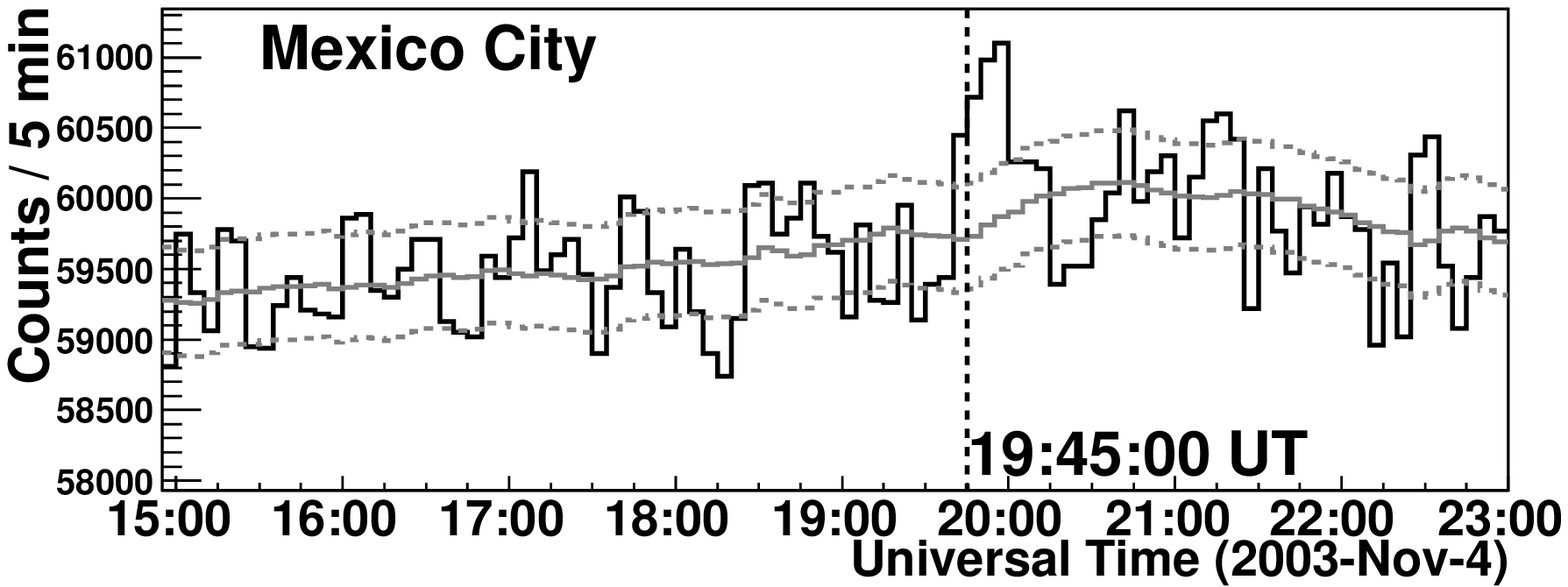}
  \caption{Five minute averages of the counting rate observed by the Haleakala (top) and Mexico City (bottom) neutron monitors on November 4, 2003. The solid smooth line is the averaged background and the dashed lines are $\pm 1 \sigma$ from the background.}
  \label{20031104_NM}
\end{figure}
\clearpage
Solar neutrons were observed by the 18NM64 neutron monitor at Haleakala, Hawaii ($203.7^{\circ}$E, $20.7^{\circ}$N, $3030\,{\rm m}$ a.s.l.).
The top panel of Figure \ref{20031104_NM} shows the five minute averages of the counting rate observed on November 4, 2003.
At this time, the sampling interval of the Haleakala neutron monitor was 10 seconds.
Clear excesses were seen after 19:45\,UT, continuing for 15 minutes.
The statistical significances of these excesses are $ 4.5 \,\sigma $ for 19:46:20 $-$ 19:51:20\,UT, $ 5.3 \,\sigma $ for 19:51:20 $-$ 19:56:20\,UT, and $ 3.1 \,\sigma $ for 19:56:20 $-$ 20:01:20\,UT.
The total significance for the fifteen minutes between 19:46:20\,UT and 20:01:20\,UT is $ 7.5 \,\sigma $.
Note that this time interval was just taken to get the maximum significance.

Solar neutrons were also observed by the 6NM64 neutron monitor at Mexico City ($260.8^{\circ}$E, $19.33^{\circ}$N, $2274\,{\rm m}$ a.s.l.), as shown in the bottom panel of Figure \ref{20031104_NM}.
At this time, the sampling interval of the Mexico City neutron monitor was 5 minutes.
Clear excesses were seen after 19:45\,UT, which was the same time as the excesses observed by the Haleakala neutron monitor.
The statistical significances of these excesses are $ 2.6 \,\sigma $ for 19:45$-$19:50\,UT, $ 3.1 \,\sigma $ for 19:50$-$19:55\,UT and $ 3.3 \,\sigma $ for 19:55$-$20:00\,UT.
The total significance for the fifteen minutes between 19:45\,UT and 20:00\,UT is $ 5.2 \,\sigma $.
\clearpage
\begin{figure}[tbp]
    \includegraphics[scale=0.8]{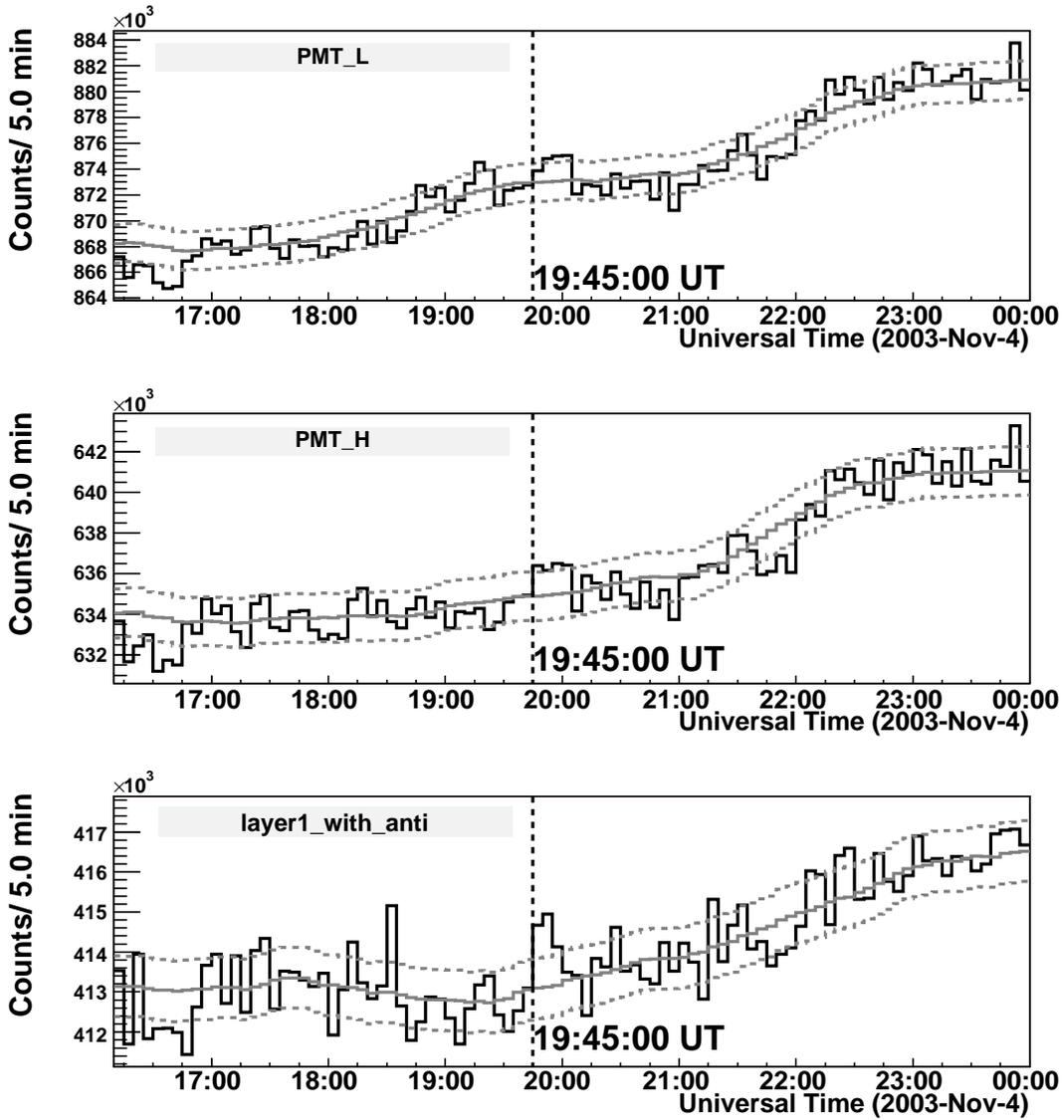}
  \caption{The five minute averages of the counting rate of PMT\_L, PMT\_H, and layer1\_with\_anti channels (see details in the text) of the Hawaii solar neutron telescope on November 4, 2003. The solid smooth line is the averaged background and the dashed lines are $\pm 1 \sigma$ from the background.}
  \label{20031104_Hawaii_SNT}
\end{figure}
\clearpage
One would expect that Mauna Kea ($203.7^{\circ}$E, $19.8^{\circ}$N, $4200\,{\rm m}$ a.s.l.) should be a better place to observe neutrons in this event than Haleakala and Mexico City.
This is the location of the Hawaii solar neutron telescope with an area of $8\,{\rm m}^2$, constructed from proportional counters and plastic scintillators, but only a minimal excess was seen after 19:45\,UT in the PMT\_L, PMT\_H, and layer1\_with\_anti channels in this telescope as shown in Figure \ref{20031104_Hawaii_SNT}.
The PMT\_L and PMT\_H are channels of scintillation counter which detect neutrons (recoil protons), energy thresholds of which are $12\,{\rm MeV}$ and $20\,{\rm MeV}$, respectively.
The layer1\_with\_anti is proportional counter channel which is located under the scintillation counters.
This apparent discrepancy between neutron monitors and the solar neutron telescope is discussed in the next section (Section \ref{sim_delta_20031104}) considering the surrounding environment of the detector.

\subsection{Analysis result}
\label{analysis_20031104}

In order to understand the Mauna Kea result, we first use the other observational data to estimate the energy spectrum of the solar neutrons.
We begin with the data from the Haleakala monitor, because it recorded the largest excess with the best time resolution.
We determine the neutron energy by using the TOF method, assuming that all the solar neutrons were produced at 19:45\,UT, the peak of the intense emission of high energy $\gamma$-rays observed by {\it INTEGRAL} as shown in Figure \ref{20031104_integral_line}.
Under this assumption, the energy of neutrons observed by the Haleakala neutron monitor between 19:51:20 and 19:56:20\,UT ranged from $59 - 913\,{\rm MeV}$.

To derive the energy spectrum of neutrons at the solar surface from the observed time profile by the neutron monitor, the survival probability of neutrons between the Sun and the Earth, the attenuation of solar neutrons passing through the Earth's atmosphere and the detection efficiency of the neutron monitor must be taken into account.
Attenuation is calculated using the Shibata model \citep{Shibata1994}, and we used the detection efficiency calculated by \citet{ClemDorman2000}.
\clearpage
\begin{figure}[tbp]
    \includegraphics[scale=0.75]{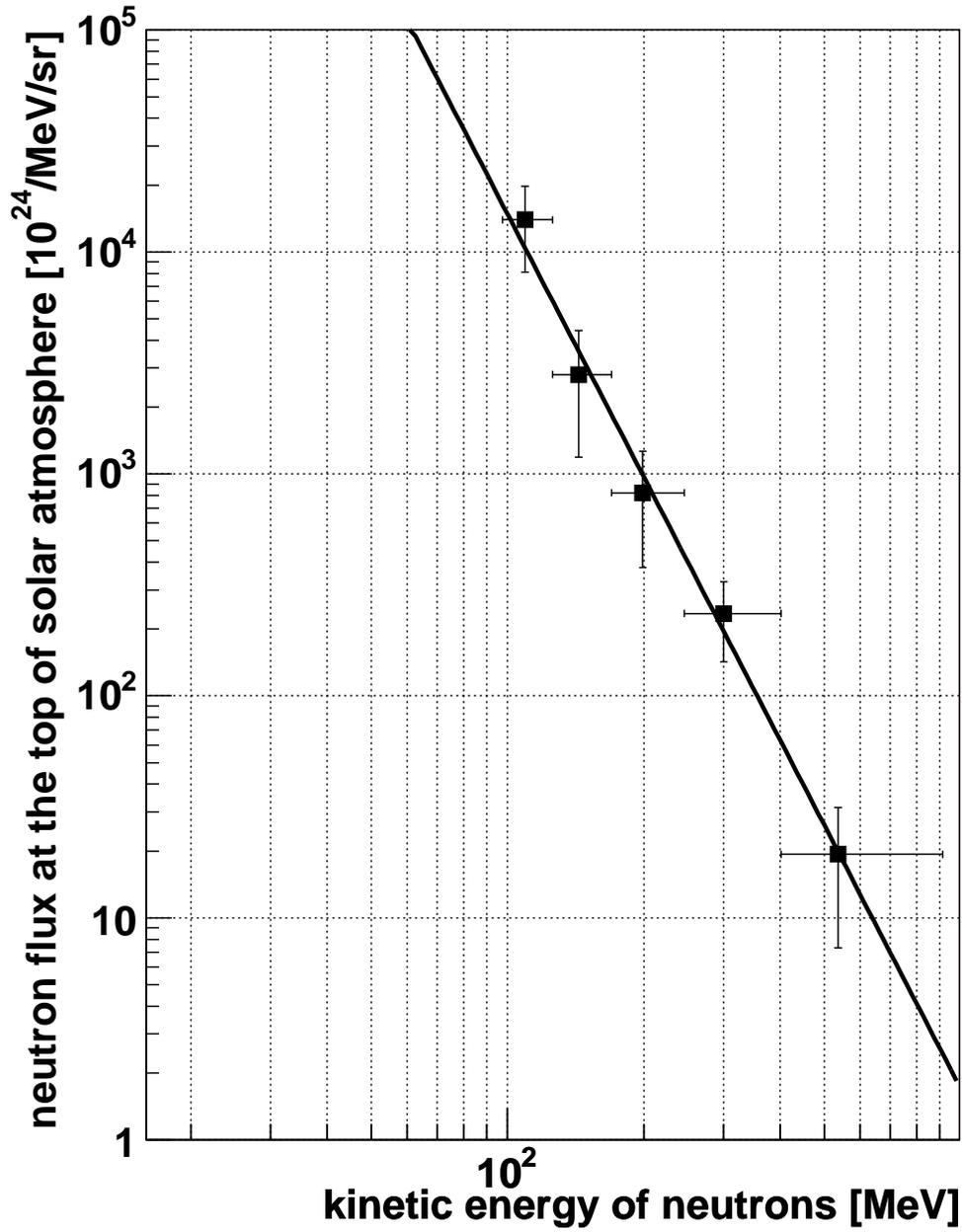}
  \caption{The energy spectrum of neutrons at the solar surface on November 4, 2003, calculated from the data of the Haleakala neutron monitor.}
  \label{20031104_energy_spec}
\end{figure}
\clearpage
Using these observational and simulation results, we calculated the energy spectrum of neutrons at the solar surface using the same method as Section \ref{observations_20031028}.
The result is shown in Figure \ref{20031104_energy_spec}.
This spectrum was derived from two minute averages of the counting rate, with the vertical errors shown only statistical.
The energy spectrum is well fitted by a power law as:
\begin{eqnarray}
Q = (1.5 \pm 0.6) \times 10^{28} \times \left( \frac{E_n}{100\,{\rm [MeV]}} \right) ^{-3.9 \pm 0.5}\,{\rm [/MeV/sr]}.
\label{eq:flux_20031104}
\end{eqnarray}
The fitting region is chosen as $ 100 {\rm \,MeV} $ and above, because there the errors from neutron attenuation in the Earth's atmosphere are small.
For this fit, $ \chi ^2 / {\rm dof} = 0.92 / 3 = 0.31 $ and the $ \chi ^2 $ probability is 82 \%.
This spectral index is typical of solar neutron events observed thus far.
The total energy flux of neutrons emitted from the Sun in the energy range $ 59 - 913\,{\rm MeV} $ is estimated to be $ 3.4 \times 10^{26}\,{\rm erg/sr} $.

\subsubsection{Simulation Using the Impulsive Model}
\label{sim_delta_20031104}
\clearpage
\begin{figure}[tbp]
    \includegraphics[scale=0.7]{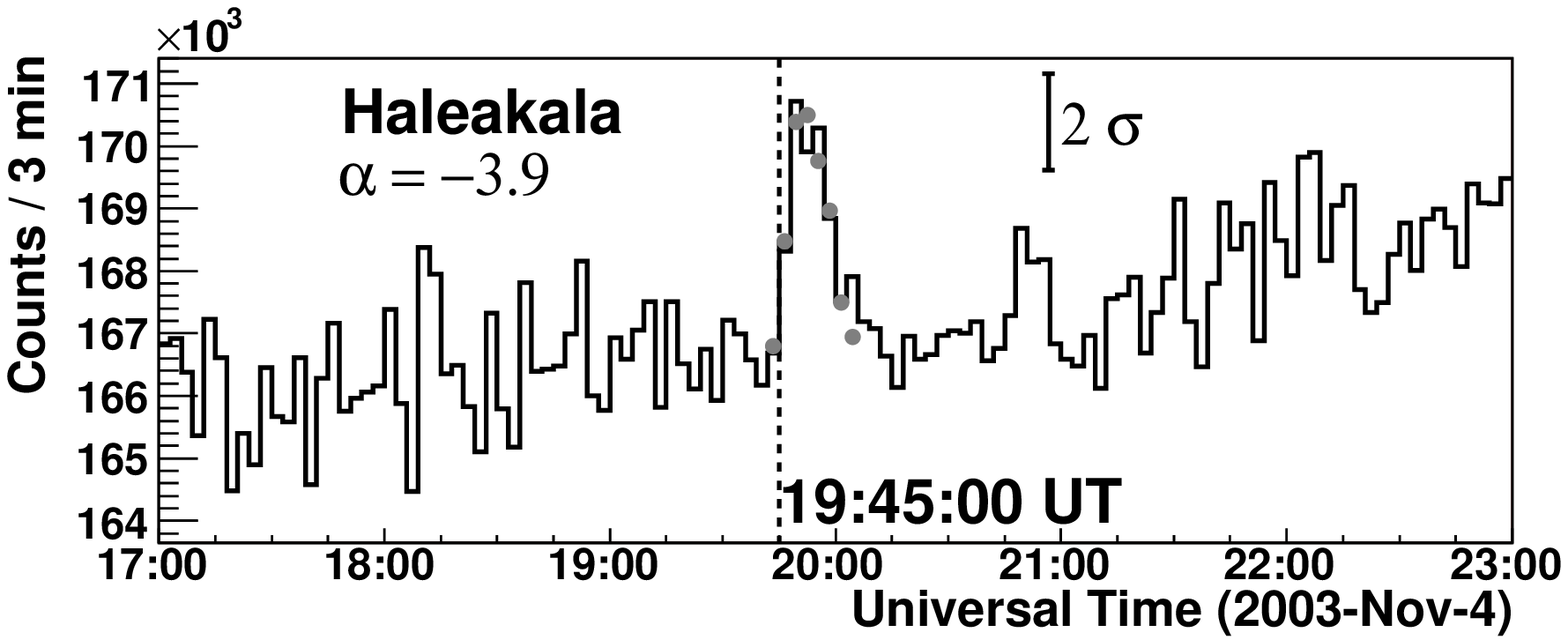}
    \includegraphics[scale=0.7]{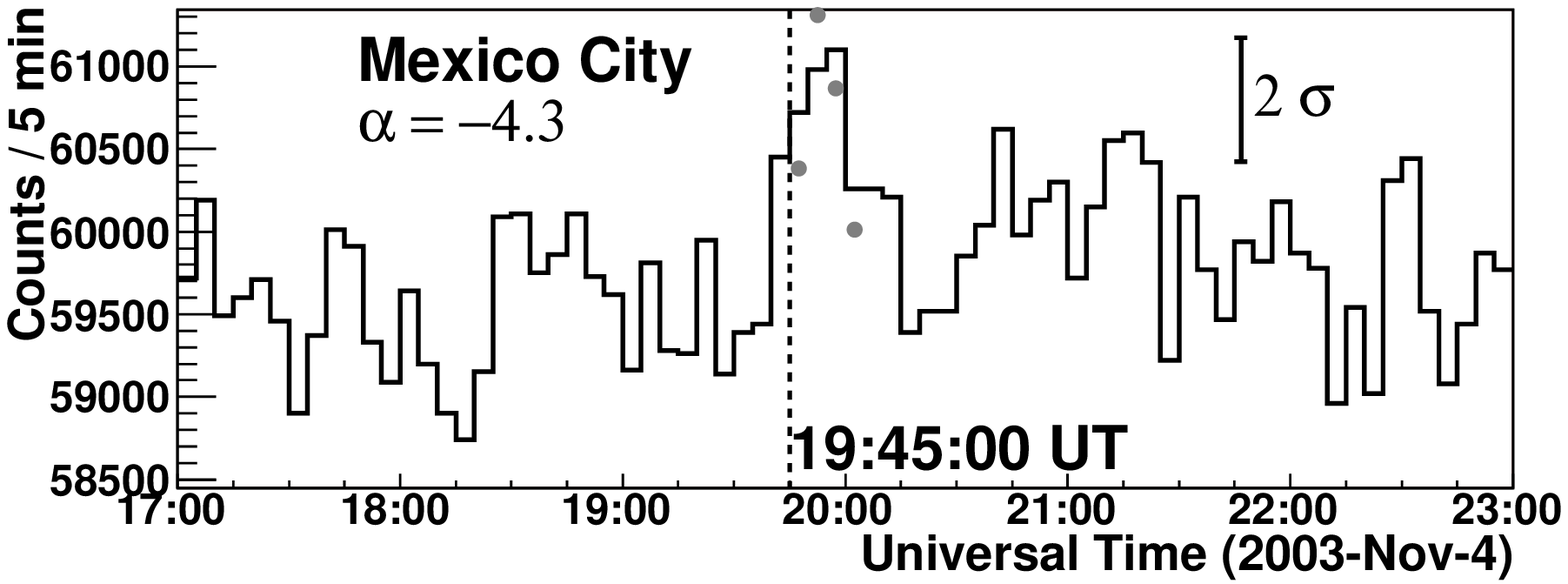}
  \caption{The best fit simulated time profiles (dots) when the spectral index is $-3.9$ for Haleakala (top), and $-4.3$ for Mexico City (bottom), superposed on the observed counting rate. The start time of the simulated time profile is 19:45\,UT, corresponding to the peak time of $\gamma$-ray emission.}
  \label{20031104_sim_tprof}
\end{figure}
\clearpage
By using the same method of Section \ref{sim_delta_20031028}, time profiles of solar neutrons assuming their spectral index at the solar surface was simulated if the neutrons were produced impulsively.
We examine the $ \chi ^2 $ of the fit of the simulated counting rate to the observed excess of the Haleakala neutron monitor obtained from Equation (\ref{eq:chi2}).
In this fitting, data obtained from 19:45 to 20:06\,UT are used.
$ \chi ^2 $ has its smallest value when the spectral index is  around $-3.9$.
When the spectral index is $-3.9$, $ \chi ^2 / {\rm dof} = 10.3 / 7 = 1.47 $, which yields the minimum $ \chi ^2 $ for the simulated time profile (the top figure of Figure \ref{20031104_sim_tprof}).
From this fitting, the energy spectrum is determined as follows:
\begin{eqnarray}
Q = (2.1 ^{+0.2} _{-0.1}) \times 10^{28} \times \left( \frac{E_n}{100\,{\rm [MeV]}} \right) ^{-3.9 ^{+0.1} _{-0.2}}\,{\rm [/MeV/sr]}.
\label{eq:flux_20031104_haleakala}
\end{eqnarray}
This is consistent to the result obtained using the simpler method shown in Equation (\ref{eq:flux_20031104}).
The total energy flux of solar neutrons with the energy range between $ 50 - 1500\,{\rm MeV} $ is $ (6.7 ^{+0.5} _{-0.4}) \times 10^{26}\,{\rm erg/sr} $, about the same order as calculated value from Equation (\ref{eq:flux_20031104}).

We have done the same analysis for the Mexico City neutron monitor.
In this fitting, data obtained during 19:45$-$20:05\,UT are used.
$ \chi ^2 $ has its smallest value when the spectral index is  around $-4.3$.
When the spectral index is $-4.3$, $ \chi ^2 / {\rm dof} = 5.55 / 3 = 1.85 $, which yields the minimum $ \chi ^2 $ for the simulated time profile (the bottom figure of Figure \ref{20031104_sim_tprof}).
From this fitting, the energy spectrum is determined as:
\begin{eqnarray}
Q = (1.6 \pm 0.2) \times 10^{28} \times \left( \frac{E_n}{100\,{\rm [MeV]}} \right) ^{-4.3 \pm 0.4}\,{\rm [/MeV/sr]}.
\label{eq:flux_20031104_mexico}
\end{eqnarray}
Although the spectral index is softer than the result of the Haleakala neutron monitor, it is consistent with Equation (\ref{eq:flux_20031104}).
The total energy flux of solar neutrons with energies between $ 50 - 1500\,{\rm MeV} $ is calculated to be $ (5.4 \pm 0.7) \times 10^{26}\,{\rm erg/sr} $, about the same order as the result obtained from the analysis of Haleakala.

We then simulated the time profile of neutrons which should be observed from the Hawaii solar neutron telescope using the energy spectrum of incident neutrons obtained from the data of the Haleakala neutron monitor, namely $ 1.5 \times 10^{28} \times (En/100\, {\rm [MeV]}) ^{-3.9} [{\rm /MeV/sr}] $.
We did not attempt to derive a spectrum from the data, because excesses of the solar neutron telescope are small.
\clearpage
\begin{figure}[tbp]
    \includegraphics[scale=0.69]{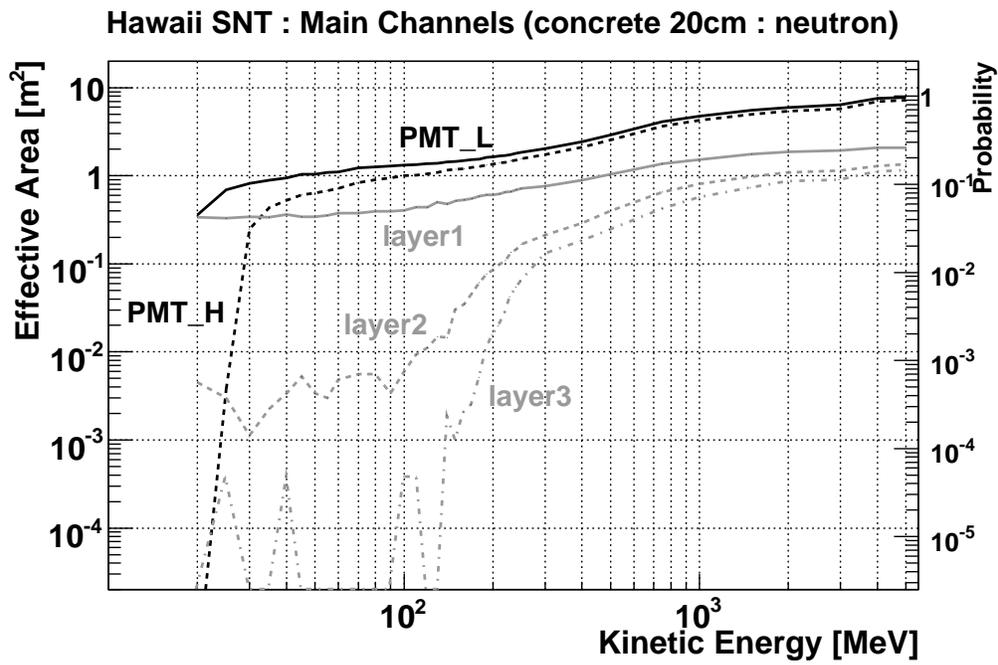}
  \caption{Detection efficiencies of the Hawaii solar neutron telescope for neutrons when the detector is surrounded by a $20\,{\rm cm}$ concrete wall. The black lines indicate the PMT\_L (solid) and PMT\_H (dashed) scintillator channels. The gray lines indicate layer channels of layer1 (solid), layer2 (dashed), and layer3 (dash-dotted) with anti coincidence of the anti counter.}
  \label{Hawaii_DetectionEfficiency}
\end{figure}
\clearpage
The detection efficiency of the Hawaii solar neutron telescope is calculated using Geant3, FLUKA$-$COLOR model.
In this calculation, the Hawaii solar neutron telescope is surrounded by the $20\,{\rm cm}$ concrete walls, since it is constructed within the building housing of the SUBARU telescope.
Figure \ref{Hawaii_DetectionEfficiency} shows the detection efficiencies of the Hawaii solar neutron telescope for neutrons and $\gamma$-rays.
\clearpage
\begin{figure}[tbp]
    \includegraphics[scale=0.75]{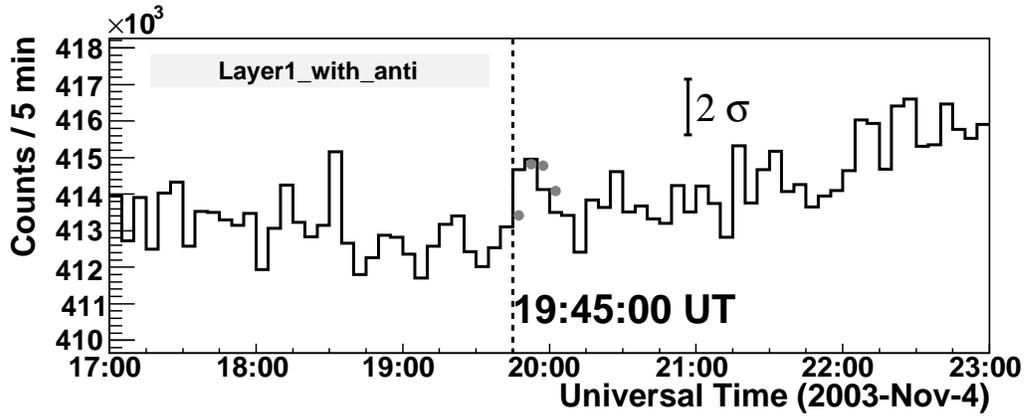}
  \caption{The simulated time profile (dots) of five minute counting rate of layer1\_with\_anti channel of the Hawaii solar neutron telescope on November 4, 2003, superposed on the observational data. The energy spectrum of the incident neutrons is $ 1.5 \times 10^{28} \times (En/100\, {\rm [MeV]}) ^{-3.9} [{\rm /MeV/sr}]$, which was obtained from the data of Haleakala neutron monitor. The start time of this time profile is 19:45\,UT, corresponding to the peak time of $\gamma$-ray emission.}
  \label{20031104_Hawaii_SNT_sim}
\end{figure}
\clearpage
The simulated result for the layer1\_with\_anti channel of the Hawaii solar neutron telescope, which recorded largest excess, is shown in Figure \ref{20031104_Hawaii_SNT_sim}.
For this fitting, $\chi ^2 / {\rm dof} = 1.24 / 3 = 0.41$, so the simulation result is consistent with the observed excess.
Because of the high counting rate from the non-hadronic component ($\gamma$-ray, muon and so on) in the Hawaii solar neutron telescope, these excesses are not statistically significant, although they correspond to the same total flux of solar neutrons observed by the Haleakala neutron monitor.
It is however possible that solar neutrons actually produced the little hump in the data.

\subsubsection{Simulation by Neutron Production Using the $\gamma$-ray Profile}
\label{sim_gamma_20031104}
\clearpage
\begin{figure}[tbp]
    \includegraphics[scale=0.7]{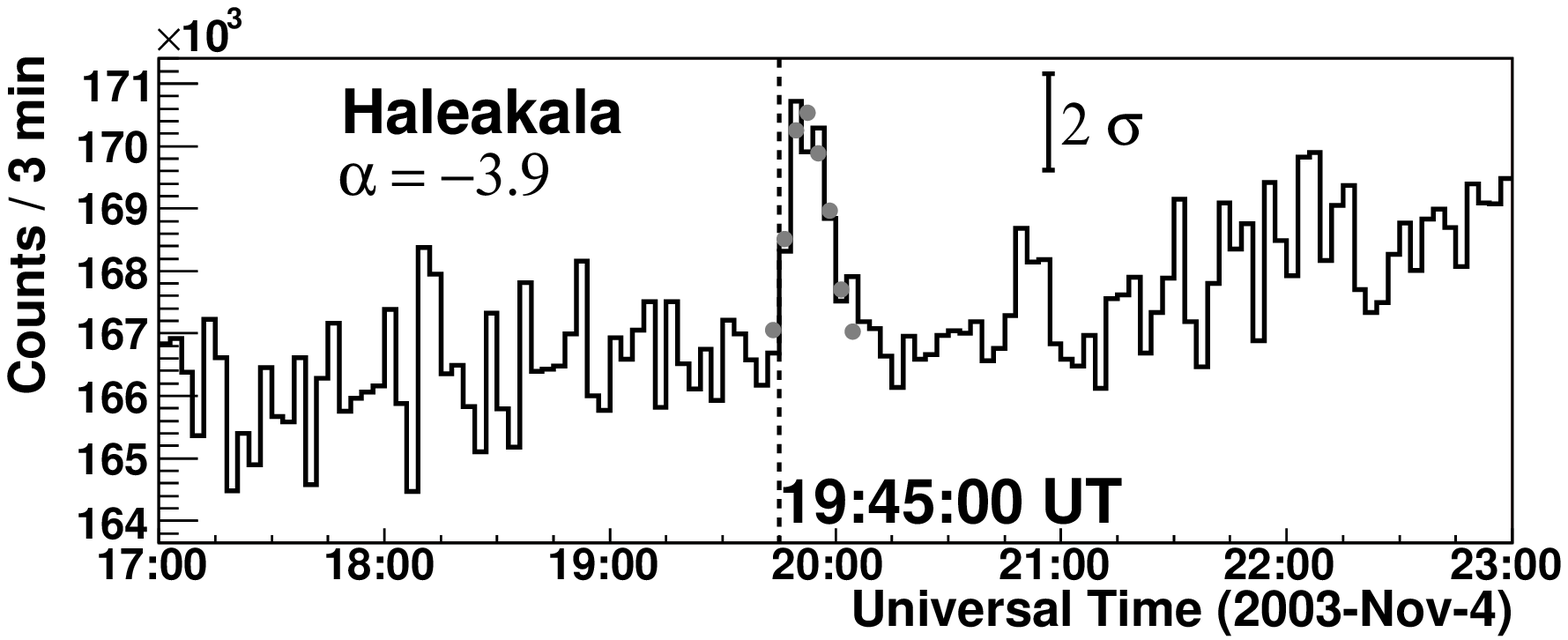}
    \includegraphics[scale=0.7]{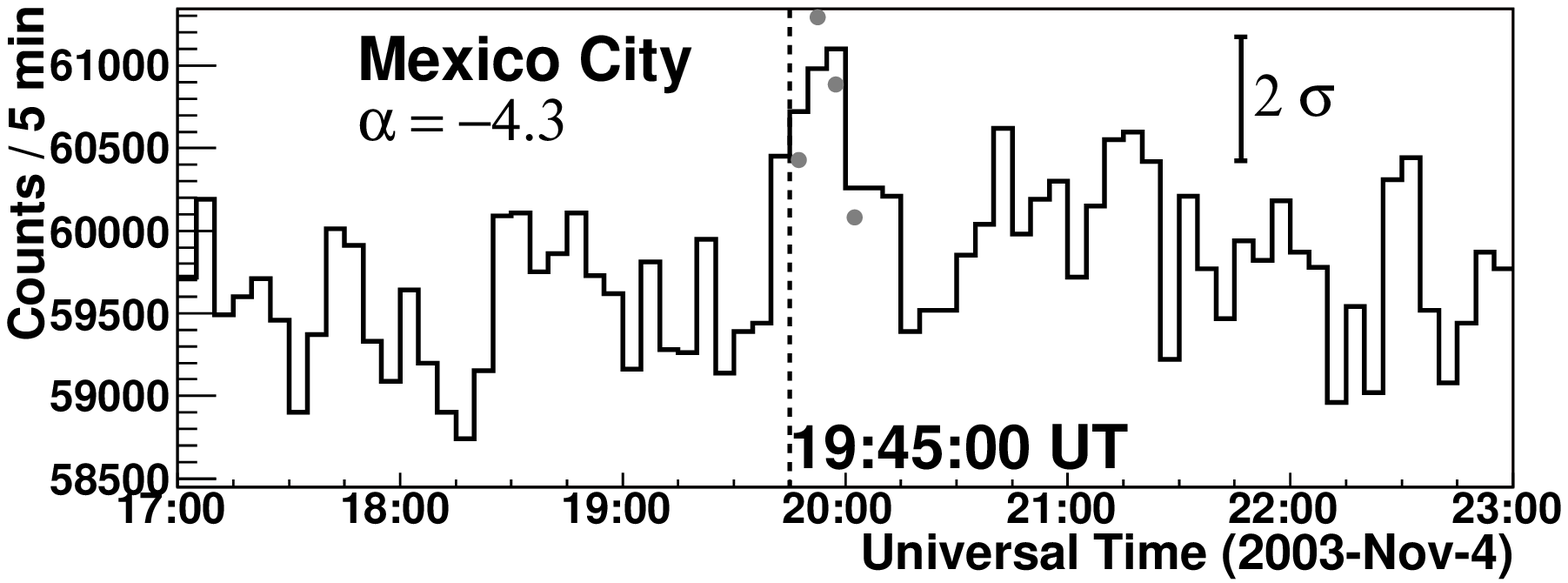}
  \caption{The simulated time profiles (dots) when the spectral index is $-3.9$ for Haleakala (top) and $-4.3$ for Mexico City (bottom), superposed on the observed counting rate of the Haleakala neutron monitor. Points are the simulated time profile for solar neutrons assuming that they were produced with the same time profile as the high energy $\gamma$-rays shown in Figure \ref{20031104_integral_line}.}
  \label{20031104_sim_tprof_gamma}
\end{figure}
\clearpage
Next we simulated the neutron time profiles detected at Haleakala and Mexico City by assuming that neutrons were produced with a time spread.
Calculated method is same as Section \ref{sim_gamma_20031028}
For this calculation, we used the $\gamma$-ray time profile observed by the {\it INTEGRAL} satellite during 19:42$-$19:48:00\,UT as the production time profile of solar neutrons, as shown in the bottom panel in Figure \ref{20031104_integral_line}.
The $ \chi ^2 $ of the fit between observed and simulated time profiles of the Haleakala and the Mexico City neutron monitor were calculated.
In this fitting, data obtained from 19:42 to 20:06\,UT are used for the Haleakala and from 19:45 to 20:05\,UT for the Mexico City neutron monitor.
For the Haleakala data, when the power index is $-3.9$ (top figure of Figure \ref{20031104_sim_tprof_gamma}), $ \chi ^2 / {\rm dof} = 10.57 / 7 = 1.51 $, giving the minimum value among the simulated time profiles.
The spectral index is determined to be $ -3.9 \pm 0.2 $.
For the Mexico City data, when the power index is $-4.3$ (bottom figure of Figure \ref{20031104_sim_tprof_gamma}), $ \chi ^2 / {\rm dof} = 4.28 / 3 = 1.43 $, giving the minimum value among the simulated time profiles and the spectral index  determined to be $ -4.3 ^{+0.4} _{-0.5} $.
The best fit spectral indices are same as those derived by assuming that the neutrons were produced impulsively.
The total energy fluxes of neutrons are estimated to be $ (7.0 \pm 0.5) \times 10^{26}\,{\rm erg/sr} $ from the Haleakala data, and $ (5.7 ^{+0.7} _{-0.8}) \times 10^{26}\,{\rm erg/sr} $ from the Mexico City data.

\section{Discussion and Summary}

Relativistic neutrons were detected in association with the X17.2 solar flare on October 28, 2003, and the X28 solar flare on November 4, 2003.
The detection of October 28 event was made by the Tsumeb neutron monitor, and the detection of November 4 event was made simultaneously by the neutron monitors at Haleakala and Mexico City, and also by the solar neutron telescope at Mauna Kea.
During these events, intense emissions of high energy $\gamma$-rays were observed by the {\it INTEGRAL} satellite.

In order to investigate the production time of solar neutrons, we compared the solar neutron data with the $\gamma$-ray data obtained from {\it INTEGRAL}.
In October 28 event, $\gamma$-ray lines from neutron capture and excited ions of C and O nuclei were clearly observed and were quite different from the time profile of bremsstrahlung $\gamma$-rays.
It appears that the time profile of electron acceleration was distinctly different from the time profile of ion acceleration.
From the time profile of neutron capture $\gamma$-rays, it seems that high energy neutrons were produced with the same time profile as $\gamma$-ray lines of de-excited ions.
In November 4 event, the time profiles of $\gamma$-ray lines which we have assumed to represent the time profile of solar neutron production, cannot be obtained independently, since the bremsstrahlung component was strong and the line $\gamma$-ray components were buried in bremsstrahlung.
However, from the time profile of the $2.2\,{\rm MeV}$ neutron capture $\gamma$-rays, it appears that the time profile of ion acceleration was approximately same as that of bremsstrahlung emissions.
Assuming that solar neutrons were produced at the time when these $\gamma$-rays were emitted, we could explain the observed excesses.

If we assume that solar neutrons were produced impulsively at 11:05\,UT on October 28, and at 19:45\,UT on November 4, when the $\gamma$-ray lines peak, we can derive the energy spectrum of solar neutrons at the solar surface from the neutron monitors as data by using Equation (\ref{eq:flux_20031028}) and (\ref{eq:flux_20031104}), respectively.
For November 4 event, in order to examine whether all excesses observed by the Haleakala and the Mexico City neutron monitor and the solar neutron telescope at Mauna Kea can be expressed by one energy spectrum consistently, and for more detailed analysis, we simulated time profiles of solar neutrons for these detectors, and compared with observed time profiles.
All of the simulation results are consistent with Equation (\ref{eq:flux_20031104}) within the range of error as shown in Equation (\ref{eq:flux_20031104_haleakala}) and (\ref{eq:flux_20031104_mexico}).
Thus, we could explain all observations with consistent spectrum.

Although we can fit the data by assuming that solar neutrons are produced impulsively, it is more natural to assume that solar neutrons are produced continuously over a finite time.
We therefore modeled the time profiles of solar neutrons by assuming that the neutrons were produced with the same time profile as $\gamma$-ray lines from excited ions. 
For October 28 event, the spectral indices thus derived are same as those derived by assuming neutrons were produced impulsively, $-3.9$ for Haleakala and $-4.3$ for Mexico City.
In other events, the spectral indices derived by assuming neutrons are produced continuously tend to be harder than those derived by assuming neutrons are produced impulsively.
However, in this event, since the $\gamma$-rays have a symmetric time profile centering around 19:45\,UT, there is little difference between the two models.
For November 4 event, the result was that the index obtained using the line $\gamma$-ray time profile is clearly harder ($-2.9$) than that obtained using the impulsive model ($-3.5$).
Therefore, for these two events, the observations were explained by assuming that solar neutrons were produced with same time profile as $\gamma$-ray lines.

Although different spectral indices are obtained with the different approach for October 28 event, these spectral indices are all consistent with the indices calculated from line $\gamma$-ray observation by the {\it RHESSI} satellite \citep{Share2004} and {\it INTEGRAL} satellite \citep{Tatischeff2005}.

The spectrum of accelerated ions can be calculated from the neutron spectrum using the spectrum of escaping neutrons produced by the accelerated ions \citep{HuaLingenfelter1987a, HuaLingenfelter1987b, Hua2002}.
From the neutron spectra shown in Equation (\ref{eq:flux_20031028}) and (\ref{eq:flux_20031104}), the number of protons above $ 30 {\rm \,MeV} $ would be about $ 10 ^{32}\,{\rm /sr} $ under the assumption that there is no turnover of the spectrum.
This is a typical value for solar neutron events observed thus far.

\acknowledgments

The authors wish to thank the {\it INTEGRAL} team, for their support to the mission and guidance in the analysis of the {\it INTEGRAL} satellite data. 
In particular, we thank Dr.\,A.\,Bykov and Dr.\,M.\,Mendez for having kindly permitted us to use their data in advance.
We acknowledge the member who is managing and maintaining the Tsumeb, Lomnicky Stit, Haleakala and Mexico City neutron monitors.
We also thank Prof.\,E.\,Fl\"{u}eckiger, Dr.\,R.\,B\"{u}tikofer of the Cosmic Ray Group of the Physikalisches Institut, University of Bern, Switzerland, members of the Armenia group, especially Prof.\,A.\,A.\,Chilingarian and Dr.\,N. Gevorgyan, and the members of the Subaru Telescope for managing and maintaining the Switzerland, Armenia and Hawaii solar neutron telescopes.
We also thank Prof. Paul Evenson for reading this manuscript.
We wish to thank the referee for evaluating this paper and for helping us to clarify several arguments.


\begin{thebibliography}{}
\bibitem[Bieber et al., 2005]{Bieber2005}
  Bieber et al., 2005, \grl, 32, L03S02.
\bibitem[Chupp et al., 1987]{Chupp1987}
  Chupp, E. L., et al., 1987, \apj, 318, 913.
\bibitem[Clem \& Dorman, 2000]{ClemDorman2000}
  Clem, J. M., \& Dorman, L. I., 2000, \ssr, 93, 1, 335
\bibitem[Debrunner et al., 1997]{Debrunner1997}
  Debrunner, H., et al., 1997, \apj, 479, 997.
\bibitem[Hua \& Lingenfelter, 1987a]{HuaLingenfelter1987a}
  Hua, X-M., \& Lingenfelter, R. E., 1987a, \apj, 323, 779.
\bibitem[Hua \& Lingenfelter, 1987b]{HuaLingenfelter1987b}
  Hua, X-M., \& Lingenfelter, R. E., 1987b, \solphys, 107, 351.
\bibitem[Hua et al., 2002]{Hua2002}
  Hua, X-M., et al., 2002, \apjs, 140, 563.
\bibitem[Lockwood \& Debrunner, 1999]{LockwoodDebrunner1999}
  Lockwood, J. A., \& Debrunner, H., 1999, \ssr, 88, 3, 483.
\bibitem[Muraki et al., 1992]{Muraki1992}
  Muraki, Y., et al., 1992, \apjl, 400, L75.
\bibitem[Muraki \& Shibata, 1996]{MurakiShibata1996}
  Muraki, Y., \& Shibata, S., 1996, AIP Conf. Proc., 374, 256.
\bibitem[Panasyuk et al., 2004]{Panasyuk2004}
  Panasyuk, M. I., et al., 2004, Cosmic Res., 42, 5, 435.
\bibitem[Share et al., 2004]{Share2004}
  Share, G. H., et al., 2004 \apjl, 615, L169.
\bibitem[Shea et al., 1991]{Shea1991}
  Shea, M.A., et al., 1991, \grl, 18, 1655.
\bibitem[Shibata et al., 1993]{Shibata1993}
  Shibata, S., et al., 1993, 23rd Inter. Cosmic Ray Conf. (Calgary), 3, 95.
\bibitem[Shibata, 1994]{Shibata1994}
  Shibata, S., 1994, \jgr, 99, A4, 6651.
\bibitem[Struminsky et al., 1994]{Struminsky1994}
  Struminsky, A., et al., 1994, \apj, 429, 400.
\bibitem[Tatischeff et al., 2005]{Tatischeff2005}
  Tatischeff, V., et al., 2005, astro-ph/0501121.
\bibitem[Tsuchiya et al., 2001]{Tsuchiya2001}
  Tsuchiya, H., et al., 2001, Nucl. Inst. Meth., A463, 183.
\bibitem[Usoskin et al., 1997]{Usoskin1997}
  Usoskin, I. G., et al., 1997, Ann. Geophysicae, 15, 375.
\bibitem[Vald\'{e}s-Galicia et al., 2000]{Valdes2000}
  Vald\'{e}s-Galicia, J. F., et al., 2000, \solphys, 191, 409.
\bibitem[Vald\'{e}s-Galicia et al., 2004]{Valdes2004}
  Vald\'{e}s-Galicia, J. F., et al., 2004, Nucl. Inst. Meth., A535, 656.
\bibitem[Veselovsky et al., 2004]{Veselovsky2004}
  Veselovsky, I. S., et al., 2004, Cosmic Res., 42, 5, 489.
\bibitem[Wang \& Ramaty, 1974]{WangRamaty1974}
  Wang, H. T., \& Ramaty, R., 1974, \solphys, 36, 129.
\bibitem[Watanabe et al., 2003]{Watanabe2003}
  Watanabe, K., et al., 2003, \apj, 592, 590.
\end{thebibliography}
\end{document}